\documentclass[apj]{emulateapj}
\usepackage{apjfonts}

\newcommand\oii{[O$\,${\sc\rmfamily ii}]\relax }%
\newcommand\oiii{[O$\,${\sc\rmfamily iii}]\relax }%

\begin{document}

\title{A Wide-field Survey of Two Z$\sim$0.5 Galaxy Clusters:
 Identifying the Physical Processes Responsible for the Observed
 Transformation of Spirals into S0s}
\author{Sean M. Moran\altaffilmark{1},
  Richard S. Ellis,\altaffilmark{1}, Tommaso Treu\altaffilmark{2,6}, Graham P. Smith\altaffilmark{3},
 R. Michael Rich\altaffilmark{4}, Ian Smail\altaffilmark{5}}
\altaffiltext{1}{California Institute of Technology, Department of
  Astronomy, Mail Code 105-24, Pasadena, CA
  91125, USA email: smm@astro.caltech.edu, rse@astro.caltech.edu}
\altaffiltext{2}{Department of Physics, University of California, Santa Barbara,
  CA 93106, email: tt@physics.ucsb.edu}
\altaffiltext{3}{School of Physics \& Astronomy, University of Birmingham,
  Edgbaston, Birmingham, B15 2TT, UK.}
\altaffiltext{4}{University of California at Los Angeles, Department of Physics \&
  Astronomy, Los Angeles, CA 90095}
\altaffiltext{5}{Institute for Computational Cosmology, Durham University, Durham
   DH1 3LE, UK}
\altaffiltext{6}{Alfred P. Sloan Research Fellow}
\slugcomment{Accepted to ApJ, July 26, 2007}
\begin{abstract}
We present new results from our comprehensive comparative survey of two
massive, intermediate redshift galaxy clusters, Cl~$0024+17$ ($z=0.39$) and
MS~$0451-03$ ($z=0.54$). Combining {\it HST} optical and 
{\it GALEX} UV imaging with Keck spectroscopy of member 
galaxies, we identify and study several key classes of `transition
objects' whose stellar populations or dynamical
states indicate a recent or ongoing change in morphology and star
formation rate. For the first time, we have been able to conclusively
identify spiral galaxies in the process of transforming into S0
galaxies. This has been accomplished by locating both spirals whose 
star formation is being
quenched as well as their eventual successors, the recently created
S0s. Differences between the two clusters in both the timescales and 
spatial location of this conversion process allow us to evaluate the
relative importance of several proposed physical mechanisms that could
be responsible for the transformation. Combined with other
diagnostics that are sensitive to either ICM-driven galaxy evolution
or galaxy-galaxy interactions--including the residuals from the 
Fundamental Plane and the properties of `signpost' compact emission
line galaxies--we describe a self-consistent picture of
galaxy evolution in clusters. We find that spiral galaxies within
infalling groups have already begun a slow process of conversion into
S0s primarily via gentle galaxy-galaxy interactions that act to quench
star formation. The fates of spirals upon reaching the core of the 
cluster depend heavily on the
cluster ICM, with rapid conversion of all remaining spirals into S0s via
ram-pressure stripping in clusters where the ICM is dense. In the
presence of a less-dense ICM, the conversion continues at a slower
pace, with galaxy-galaxy interactions continuing to play a role along
with `starvation' by the ICM. We conclude that the buildup of the
local S0 population through the transformation of spiral galaxies is a
heterogeneous process that nevertheless proceeds robustly across a
variety of different environments from cluster outskirts to cores.

\end{abstract}

\keywords{galaxies: clusters: individual (Cl 0024+1654, MS
0451-0305)
--- galaxies: elliptical and lenticular, cD --- galaxies:
evolution
--- galaxies: spiral --- galaxies: stellar content
--- ultraviolet: galaxies}

\section{INTRODUCTION}

It is well-known that environmental processes play a 
significant role in shaping the evolution of galaxies as they assemble
onto clusters. With the aid of Hubble 
Space Telescope ({\it HST}) imaging and deep optical spectroscopy, 
recent studies have quantified this evolution in galaxy properties, 
painting a picture where the fraction of early-type 
(elliptical and S0) galaxies and the fraction of passive
non-star-forming galaxies both grow with time, and at a rate that
seems to depend sensitively on the local density of galaxies
\citep{d97,poggianti99, smith05b,postman05}.

Yet there are a wide variety of physical processes that may be
responsible for these evolutionary trends--including galaxy mergers, 
galaxy-galaxy harassment, gas stripping by the ICM, or tidal processes 
\citep{moore99,fujita98, bekki02}. 
Observationally, it has so far been impossible
to fully separate the effects of the various physical processes, 
in large part due to the overlapping regimes 
of influence for each of the proposed mechanisms \citep[see][]{tt03}.
Further complicating the picture, the large scale
assembly states of clusters show considerable variety
\citep{smith05a}, such that the dominant forces acting on galaxies are 
likely to vary from cluster to cluster, or over the course of an 
individual cluster's assembly history.  
But gaining an understanding of the complex interplay between a
variable ICM, the properties of assembling galaxies, and the overall 
cluster dynamical state is crucial if we are to have a complete picture of the
growth and evolution of galaxies in a hierarchical universe.

In this paper, we combine optical ({\it HST}) and UV
({\it GALEX}) imaging of two $z\sim0.5$ galaxy clusters with 
ground-based spectroscopy of member galaxies, in an attempt to
trace directly the buildup of passive early-type galaxies via a
detailed `case study' of the galaxy population across each cluster. 
The two studied clusters, Cl~0024+17 ($z=0.40$) and MS~0451
($z=0.54$), are part of a long-term campaign to trace the 
evolution of galaxies in wide fields ($\sim10$~Mpc diameter)
centered on both clusters, using a variety of methods.
By undertaking an in-depth, wide-field comparative study of two prominent
clusters, we hope to provide a complement to other observational 
\citep[e.g.,][]{cooper07, poggianti06} and theoretical investigations 
\citep[e.g.,][]{delucia06} 
which trace with a broad brush the evolution in star formation rate and the 
buildup of structure in the universe. 

The first paper in our series, \citet{tt03} (hereafter Paper I), introduced our panoramic
{\it HST} imaging of Cl~0024 and began our ongoing discussion of the
physical processes that may be acting on galaxies within clusters.
In several subsequent papers, whose results are summarized in \S~2, we
have added extensive optical spectroscopy to the program,
allowing targeted investigations of galaxy stellar populations and
star formation rates as a function of cluster-centric radius, local 
density, and morphology. Our goal for this paper is to bring our
complete survey data set to bear on the question of how galaxies are
affected by their environment, as a function of both the overall
cluster properties and of local environment within each cluster. 
For maximum clarity and deductive power, we focus 
our investigation on several key populations of `transition
galaxies' in the clusters--galaxies whose stellar populations or dynamical
states indicate a recent or ongoing change in morphology or star
formation rate.

\begin{deluxetable*}{lcclcccc}
  \centering
  \tablewidth{0pt}
  \tablecaption{\label{tab:basic}Basic Properties of the Clusters}
  \tabletypesize{\footnotesize}
 \tablehead{\colhead{Name } & \colhead{RA} &\colhead{DEC} &\colhead{$R_{VIR}$} &
  \colhead{$M_{200}$} & \colhead{z} & \colhead{$L_X$}& \colhead{$T_X$}\\
\colhead{ } & \colhead{($^\circ$)} & \colhead{($^\circ$)} & \colhead{(Mpc)} & \colhead{($M_\odot$)} &
\colhead{} & \colhead{($L_\odot$)} & \colhead{(kev)} }
  \startdata
Cl~0024 & 6.65125 & 17.162778 &$1.7^{(1)}$ & $8.7\times 10^{14} \ ^{(2)} $ & 0.395 & $7.6 \times
10^{10} \ ^{(3)}$ & 3.5$^{(3)}$  \\
MS~0451 & 73.545417 & -3.018611 &2.6  & $1.4\times 10^{15} \  ^{(4)}$ & 0.540 & $5.3 \times
10^{11} \  ^{(4)}$ &  10.0$^{(4)}$ \\

\enddata
\tablecomments{$^{(1)}$ Treu et al. (2003), $^{(2)}$ Kneib et al. (2003), $^{(3)}$
  Zhang et al. (2005), $^{(4)}$ Donahue et al. (2003).}
\end{deluxetable*}

In evaluating cluster galaxies for signs of evolution, 
we have adopted a strategy to make maximal use of our {\it
  HST}-based morphologies by characterizing signs of recent evolution in
spirals and early-types separately. This approach is similar to
using the color-magnitude relation to divide our sample into `red
sequence' and `blue cloud' galaxies, but it provides additional
leverage to identify galaxies in transition. Early-type galaxies that
have either been newly transformed or prodded back into an 
active phase, or spiral galaxies where star formation is being 
suppressed or enhanced will all stand out in our sample. 
At the same time, their morphologies reveal 
important information about their formation histories prior to 
their current transition state, information that colors alone do not 
provide. Our strategy also has the benefit of allowing us to 
directly investigate the hypothesis that many 
cluster spirals transform into S0s between $z\sim0.5$ and today 
\citep{d97}--an investigation that will form the basis of this
paper.

In the next section, we outline our rationale for selecting Cl~0024
and MS~0451, describe the large-scale properties of each cluster, 
and give a summary of what we have concluded so far in
our study of galaxy evolution in both clusters. In \S3, we describe
new data not covered in previous papers in our series. In \S4, we
will investigate the properties of `passive spirals' across the two
clusters, suggesting that they are in the
process of transforming into S0 galaxies. We confirm in \S5
that this is the case, via identification of newly created S0s that
we believe reflect the distinct passive spiral populations 
found in each cluster.  
In \S6, we consider the environments of these galaxies in transition,
and begin to investigate the physical mechanisms that may be
responsible for these transformations.
In \S7, we outline a model of how galaxy evolution proceeds in each
cluster. We consider the 
Fundamental Plane as a way to further 
constrain the physical mechanisms at work, and derive 
similar constraints from the populations of compact 
emission line galaxies in both clusters. 
Finally, in \S8,  we summarize our conclusions about the
transformation of spirals into S0s at $z\sim0.5$.
In this paper, we adopt a standard $\Lambda$CDM cosmology with
$H_0=70$~km~s$^{-1}$~Mpc$^{-1}$, $\Omega_m=0.3$, and $\Omega_\Lambda=0.7$.

\section{A Comparative Survey of Two $z\sim0.5$ Clusters}

Cl~0024 and MS~0451 were chosen for study primarily because of their 
comparable total masses, similar galaxy richness, and strong lensing
features, while exhibiting X-ray properties that are quite distinct
(See Table~\ref{tab:basic}). While MS~0451 is one of the most X-ray
luminous clusters known \citep{donahue03}, Cl~0024 is somewhat
under-luminous in the X-ray, with a mass inferred from {\it XMM/Newton} observations
that significantly underestimates the mass derived from other methods 
\citep{zhang05}. MS~0451 has X-ray luminosity seven times larger than
Cl~0024, with a corresponding gas temperature nearly three times as high.
This implies a large difference in the density and
radial extent of the intracluster medium (ICM) between the two
clusters. As a result, ICM-related physical processes are naively
expected to be more important in the evolution of currently infalling 
MS~0451 galaxies than in Cl~0024. 

\begin{figure}
\centering
\includegraphics[width=\columnwidth]{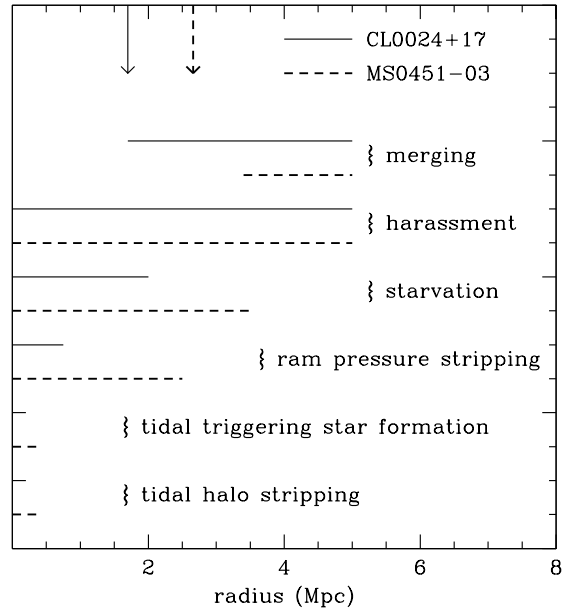}
\caption{\label{mechanisms} Schematic diagram indicating the
  cluster-centric radius over which each of several listed physical
  mechanisms may be effective at fully halting star formation or
  transforming the visual morphology of a radially infalling
  galaxy. Each physical mechanism listed can act effectively over 
  the radial range indicated by the solid (Cl~0024) or dashed line
  (MS~0451). Arrows indicate the virial radius for
  each cluster. Note that tidal processes here refer to interactions with
  the cluster potential, while tidal forces during 
  galaxy--galaxy interactions are a component of the harassment mechanism. 
  Each regime of influence was calculated according to the
  simple models described in Paper I for Cl~0024, using the
  global cluster properties given in Table~\ref{tab:basic}.}
\end{figure}

In the schematic diagram shown in Figure~\ref{mechanisms}, we apply
simple scaling relations to estimate the regimes of influence for
several key physical processes which could be acting on
infalling galaxies, following the procedure described in
Paper I. ICM-related processes, such as gas starvation
\citep{larson80, bekki02} and ram-pressure stripping \citep{gunn72} begin to affect
galaxies at much larger radius in MS~0451. As in Paper I, we expect
that other ICM-related processes such as thermal evaporation
\citep{cowie77} or viscous stripping \citep{nulson82} operate 
with at most similar strength to ram
pressure, and so we do not consider them separately.
An important caveat, however, is that the role of
difficult-to-observe shocks in the ICM are unknown, and are not
accounted for in Figure~\ref{mechanisms} (but see \S7).
Similarly, the two clusters' differing masses set the radial regions
where galaxy merging will be effective; because of the $\sim50\%$ higher
mass of MS~0451, typical galaxy relative velocities become too fast for 
mergers to occur at a higher radius than in Cl~0024.

The differing regimes of influence for the physical mechanisms
illustrated in Figure~\ref{mechanisms} provide the key template for
our attempt to disentangle the relative importance of the various
processes. By surveying the galaxies of both clusters for 
signs of recent transformation or disturbance, across the entire
radial range to $\sim5$~Mpc, we hope to associate the sites and
characteristics of galaxies in transition with the likely causes 
from Figure~\ref{mechanisms}.

An additional factor not reflected in Figure~\ref{mechanisms},
however, is the overall assembly state and level of substructure in
each cluster. Therefore, Figure~\ref{mechanisms} can only be used as
a guide, and we must carefully consider the effects that large-scale
cluster assembly and irregularities may have on their galaxies as
well. For example, the effects of tidal processes and galaxy-galaxy harassment,
which according to Figure~\ref{mechanisms} should occur in much the
same regions across both clusters, may very well differ greatly
between clusters depending on how well each is virialized. 
As we will discuss below, there are marked differences in the
levels of substructure between the two clusters, and these may drive
important differences between the galaxy populations.

\subsection{Kinematic Structure of the Two Clusters}

\begin{figure}
\centering
\includegraphics[width=\columnwidth]{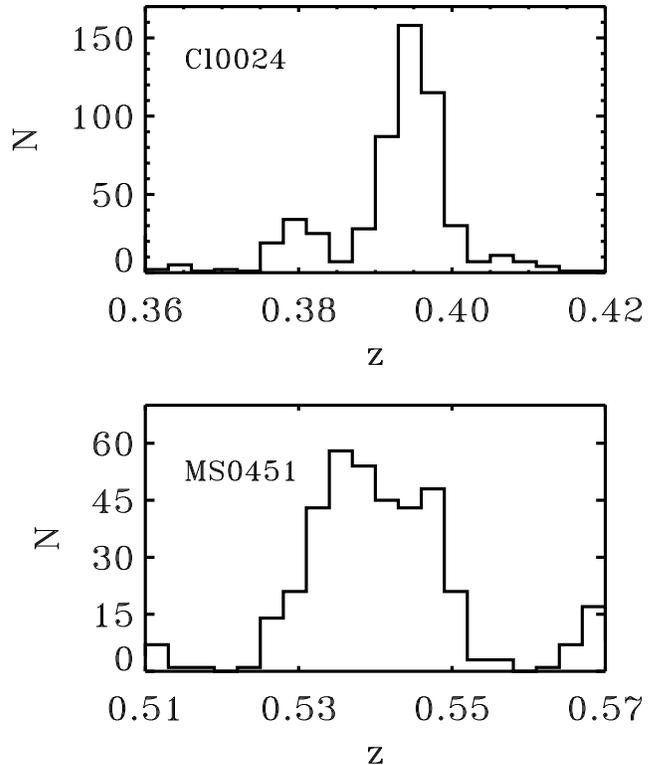}
\caption{\label{redshifts} Distribution of redshifts for both Cl~0024
  (top) and MS0451 (bottom), including all galaxies with 
  spectroscopically confirmed redshifts from our observations as well
  as previously published redshifts (see \S3). In Cl~0024, we identify
  508 cluster members, defined to lie in the range $0.373<z<0.402$. 
  In MS~0451, we count 319 cluster members in the range $0.52<z<0.56$.}
\end{figure}

While it is evident from their different X-ray luminosities that 
MS~0451 and Cl~0024 provide quite different environments for their
constituent galaxies, our detailed study of environmental
effects on cluster galaxies requires a comprehensive characterization of 
the two clusters and their respective environments. Here, 
we study the radial velocities and spatial distributions of
galaxies in each cluster, in order to evaluate the 
global kinematic structure of each cluster and identify significant 
substructures. 

Our extensive {\it HST} imaging and Keck spectroscopy, which will be
described more fully in \S3, readily reveal
marked differences in the distributions of
galaxies between the two clusters, suggesting that the clusters have quite
dissimilar recent assembly histories. In Figure~\ref{redshifts}, we
display the distribution of redshifts for members of both Cl~0024 and
MS~0451. The redshift distribution of MS~0451 members is
broadly consistent with a Gaussian distribution. A somewhat better fit
to the data is given by a two-component double Gaussian function,
but the distribution splits into these two peaks only for galaxies 
at large radius ($R>2$~Mpc), suggesting that we are observing 
two filaments feeding galaxies into the virialized center of MS~0451.   
Cl~0024 galaxies likewise exhibit a double-peaked structure in redshift
space, with two components that are widely separated and
asymmetric in height. This feature was
discovered by \citet{czoske01}, and is thought to be the remnant of a
high-speed, face-on encounter between the main cluster and a smaller
subcluster or large group \citep{czoske02}.

\begin{figure*}
\centering
\includegraphics[width=\columnwidth]{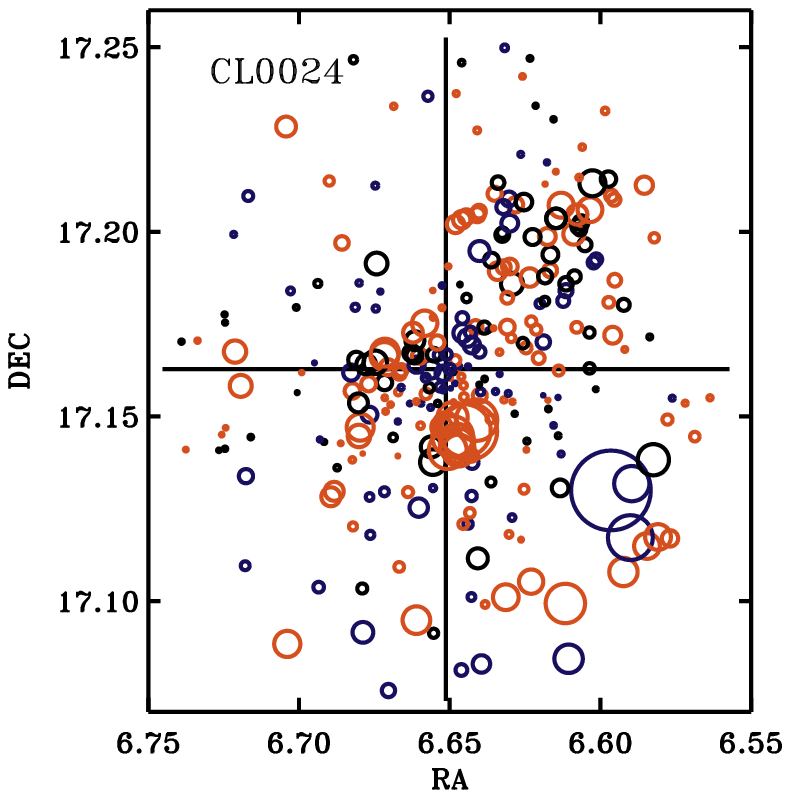}
\includegraphics[width=\columnwidth]{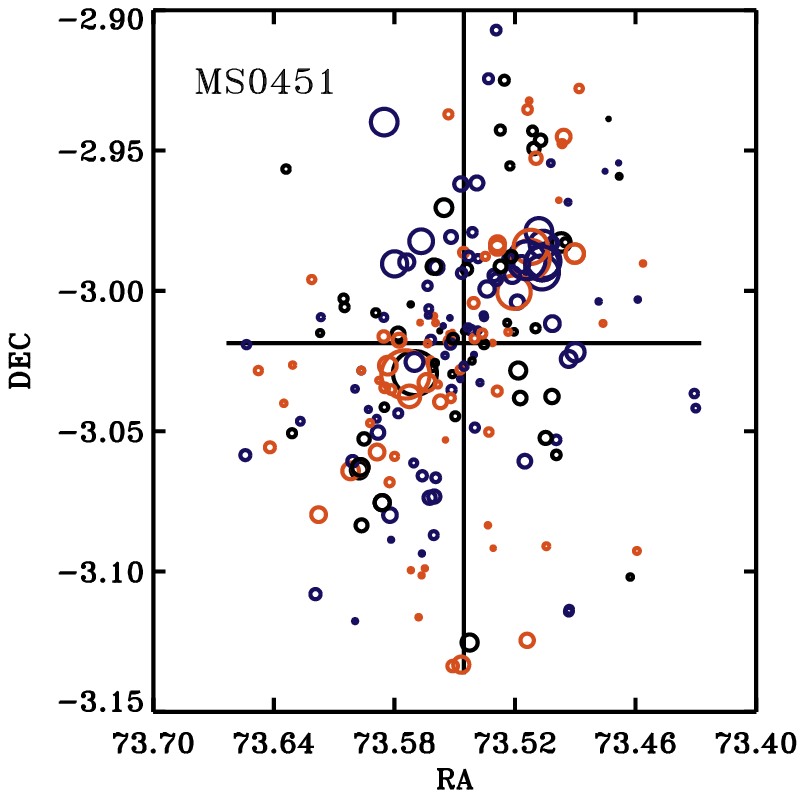}
\caption{\label{dstests}Dressler-Shectman plots for Cl0024 left, and
  MS0451, right. Including only spectroscopically confirmed objects
  within 1 R$_{V}$ of the cluster center, each circle indicates the
  spatial position of a cluster member, with the circle size 
  proportional to its local deviation from a smooth velocity 
  distribution (see text). Circles
  are coded blue for galaxies with cluster-centric velocity toward the
  observer with amplitude $>300$~km~s$^{-1}$, red for galaxies with
  the same velocity amplitude away from the observer, and black for
  galaxies with low cluster-centric velocities, $<300$~km~s$^{-1}$. }
\end{figure*}

The effects of this collision, even an estimated several Gyr after it
occurred \citep{czoske02}, are still important in the 
core of Cl~0024--for example, as shown by intriguing recent
claims of a `dark matter ring' in the core of Cl~0024 \citep{jee07}. 
The measured line of sight velocity dispersion in the core
of $650\pm50$~km~s$^{-1}$ implies a 30--50$\%$ lower mass than has been
directly measured through strong and weak lensing constraints
\citep{kneib03}. The fact that the galaxy velocity distribution and X-ray
emitting gas both underestimate the mass leads to the conclusion 
that the core of Cl~0024 is not in virial nor hydrostatic equilibrium, an
assumption that was made for each of these mass estimates.
The lower relative velocities of galaxies
in the Cl~0024 core may importantly affect the action of physical
processes whose strengths vary with galaxy velocity. For example, gas
stripping by the ICM may be even less effective in Cl~0024 than naively
predicted by Figure~1. The effects of galaxy--galaxy harassment may
also be different, as close galaxy encounters may happen at both lower
frequency and lower relative velocities than have previously been
modeled in detail \citep[e.g.,][]{moore99}.

The spatial distribution of galaxies in MS~0451 and Cl~0024 also show
key differences. In Figure~\ref{dstests}, we construct
modified Dressler-Shectman (D-S) plots for the region within $R_V$ 
of each cluster \citep{ds88}. In such plots, each cluster member is
indicated by a circle, with the size of the circle proportional to
that galaxy's `Dressler-Shectman statistic':
\begin{equation}
\delta^2=11/\sigma^2[(\bar{v}_{local}-\bar{v})^2+(\sigma_{local}-\sigma)^2]
\end{equation}
where line-of-sight velocity $\bar{v}_{local}$ and dispersion
$\sigma_{local}$ are measured with respect to each galaxy's ten
nearest neighbors, and $\bar{v}$ and $\sigma$ are the global cluster values.
This statistic measures
each galaxy's local deviation from a smooth, virialized velocity and spatial
distribution. In other words, groups of large circles on the plot tend to indicate the
presence of an infalling group. In Figure~\ref{dstests}, we further
color-code each galaxy according to its velocity with respect to the
cluster center.

The D-S plot for Cl~0024 reveals the presence of at
least two significant groups near to the cluster core: an infalling
group at high velocity nearly along the line of sight to the cluster
center, and a previously noted large group to the northwest of the
cluster core, presumably assembling onto the core in an orientation
almost in the plane of the sky. The large structure to the NW is also
detected in both the weak lensing map \citep[][hereafter, Paper
  II]{kneib03} 
and {\it XMM/Newton}
observations \citep{zhang05}, which additionally suggest the presence of a
shock front at the interface between the group and the main cluster.

In contrast, the D-S plot for MS~0451
reveals an elongated but largely smooth distribution of galaxies. The
cluster's elliptical shape is clearly visible, but the segregation of
red points to the southeast and blue points to the northwest suggests
a uniform contraction of the cluster along its major axis. While it is
possible that the observed velocity field is due to cluster-scale
rotation, previous analyses of the X-ray and Sunyaev-Zeldovich 
observations prefer a prolate or triaxial shape for MS~0451 
\citep{donahue03, defilippis05}, whereas the cluster
would necessarily be oblate if the observed velocities are due 
to rotation.  We may be
seeing this spread in velocities reflected in the redshift 
distribution of MS~0451, which splits into two peaks at large radius,
likely indicating infall from two directions. One might worry
that the elongated distribution gives MS~0451 an artificially large
velocity dispersion, $\sigma$, but cluster mass estimates derived
from $\sigma$ are consistent with those derived from the X-ray
and weak lensing analysis \citep[][and references
  therein]{donahue03}. The origin
of the elliptical shape is unclear, but may represent the remnant of
a past major merger. Nevertheless, still-bound infalling groups 
appear to be absent or insignificant within the virial radius.

The overall increased level of substructure observed in Cl~0024 is
likely to have implications for the galaxy population. In comparison to
MS~0451, Cl~0024 galaxies are more likely to have been a
member of an infalling group in the recent past, and so we may
expect to see more signs of recent group `pre-processing' in Cl~0024.
Furthermore, in the chaotic environment of Cl~0024 the ICM is
more likely to be disturbed, and so shocks, cold fronts, or other
features in the ICM may be present, and could play an important role in
the star formation histories of cluster members \citep{roettiger96}.

\subsection{Previous Work}
Between the extremely dense ICM in MS~0451 and the active assembly
state of Cl~0024, these two clusters provide very distinct
environments for their member galaxies. In the course of this paper, we
will examine how the properties of transition galaxies in each cluster 
reflect these distinct environments. However, as our new results rely on and
incorporate the findings of our previous papers in this series, 
we present here a brief summary of our investigations so far. We
highlight in particular several results which, in the initial
interpretation, hint at the action of one or more 
physical mechanisms from Figure~\ref{mechanisms}.

\begin{list}{$\bullet$}{}
\item By constructing the Fundamental Plane (FP) of Cl~0024, we
   observed in \citet{moran05} (hereafter Paper III) that
   elliptical and S0 galaxies (E+S0s) exhibit a 
   high scatter in their FP residuals, equivalent
   to a spread of $40\%$ in mass to light ratio ($M/L_V$).
   The high scatter occurs only among galaxies in the cluster core, 
  suggesting a turbulent assembly history for cluster early types, perhaps
  related to the recent cluster--subcluster merger \citep{czoske02}.

\item Around the virial radius of Cl~0024, we observed a number of
   compact, intermediate--mass ellipticals
   undergoing a burst of star formation or weak AGN activity, 
   indicated by strong \oii \ emission (Paper III). 
   The \oii \ emitters reside in relatively low density
   and high speed regions, and so we deemed
   that they are not likely the remnants of mergers. Though we will
   revisit the merger hypothesis in \S7, we tentatively concluded that 
   the observed activity is caused by a rapidly acting physical process: two 
   candidates are galaxy harassment and shocks in the ICM, perhaps 
   generated by the sub-cluster merger in Cl~0024.

\item We searched for disruptions in the internal structures and star
  formation rates of disk galaxies due to tidal effects or
  galaxy--galaxy interactions, by measuring emission line rotation curves and
  constructing the Tully-Fisher relation \citep{moran07a}.
  We find that the cluster TF relation exhibits
significantly higher scatter than the field relation, echoing the high
  scatter seen in the Cl~0024 FP.
  We argue that the high scatter in both $K$- and $V$-band TF relations 
demonstrate that cluster
spirals are kinematically disturbed by their environment. We proposed that such
disturbances may be due to galaxy merging and/or harassment.

\item We combined {\it GALEX} UV observations with key spectral line indices
  to place strict constraints on the recent star formation histories
  of ``passive spiral'' galaxies, an important class of transition
  object \citep{moran06}.
  Passive spirals show spiral morphology in {\it HST} images, 
   but reveal weak or no \oii \ 
   emission in their spectra, suggesting a lack of current star
   formation.  Through {\it GALEX} UV imaging, we find that passive 
   spirals in Cl~0024 exhibit UV emission nearly as strong as regular 
   star-forming spirals, implying the presence of young stars. Their unusual 
combination of UV emission with weak H$\delta$ strength 
supports a picture where passive spirals
have experienced a rapid decline and eventual
cessation of star formation over the last $\sim0.5-2$~Gyr.
The timescale of this decline suggests ``starvation'' by the ICM as
a possible cause \citep{larson80, bekki02}--a process where the
intracluster medium (ICM) strips gas from a galaxy's halo, causing
star formation to decline due to the absence of new cold gas accretion 
onto the disk.
\end{list}

Each of these previous investigations has revealed a partial view of
environmental evolution across the studied clusters. Through these 
investigations, we have identified several physical mechanisms that 
we could call `likely suspects' for driving galaxy evolution in
clusters, and these seem to fall into two classes: galaxy--ICM
interactions and galaxy--galaxy interactions. 
However, a unified picture is still lacking. Both of these flavors 
of interaction have been implicated before in the decline of star
formation and the possible conversion of spirals into S0s
\citep[see][and references therein]{boselli06}, yet a
detailed evaluation of their importance remains elusive. 

We aim in this paper to complete our accounting of galaxy evolution across 
these two quite distinct clusters, building on and linking our 
earlier work into what we hope will be a more comprehensive 
picture of how cluster galaxies are affected by their environment at
intermediate redshift.
To accomplish this, we will first document what we believe to be direct
evidence for the transformation of spirals into S0s: through an
analysis of their stellar populations and recent star formation rates (SFRs), 
we link the passive spiral galaxies in both clusters to their eventual end
states as newly-generated cluster S0 galaxies. Only
then, as we examine the physical mechanisms responsible for this
transformation, will we draw on the above summarized results to place
extra constraints on the physics driving the transformation. This
discussion will include an extension of our 
analysis of the Fundamental Plane and the strongly-emitting compact 
E+S0s in Cl~0024 to include MS~0451, in order to further strengthen the
constraints we can place on galaxy--galaxy interactions and
galaxy--ICM interactions, respectively.

\section{NEW DATA}
Before turning to our new analysis, we must specify the
characteristics of our now-complete data set.
While most imaging and spectroscopic data on Cl~0024 has
been fully or partially described in previous papers \citep[][Paper I,
  Paper III]{moran06}, much of the following relies on new data for
MS~0451. In this section, we describe our new observations of MS~0451, and
provide updated information on Cl~0024, including {\it GALEX} imaging and
additional spectroscopy. 

\subsection{Imaging}
We make use of {\it HST} imaging of Cl~0024
and MS~0451 from the comprehensive wide-field survey described in
Paper I and Smith et al. (2007, in
preparation). In Cl~0024, {\it HST} coverage consists of a sparsely-sampled
mosaic of 39 WFPC2 images taken in the F814W filter ($\sim I$ band),
providing coverage to a projected radius $> 5$ Mpc at exposure
times of 4-4.4ks. MS~0451
observations were taken with the ACS, also in F814W, and 
provide contiguous coverage
within a 10Mpc$\times$10Mpc box centered on the cluster, with single
orbit (2ks) depth over the field. 

For both clusters, reliable morphological classification is possible to 
rest frame absolute magnitude $M_V=-19.5$, corresponding to $I=22.1$ in
MS~0451 and $I=21.2$ in Cl~0024. 
Broader classification as early or late type
is possible to a fainter limit, $M_V=-18.0$. All galaxies that have
spectroscopically confirmed redshifts and are brighter
than this limit are classified visually following the procedure
described in Paper I for Cl~0024. In MS~0451, galaxies were 
classed by one of us (RSE), and we expect that the typing is accurate
to the quoted limits based on previous experience with ACS
imaging of similar depth \citep[e.g.,][]{tt05b}. Morphologies were assigned
according to the Medium Deep Survey scheme introduced by
\citet{abraham96}: T=-2=star, -1=compact, 0=E, 1=E/S0, 2=S0, 3=Sa+b,
4=S, 5=Sc+d, 6=Irr, 7=Unclass, 8=Merger, 9=Fault. In the following,
all galaxies assigned types T=0,1,2 are together labeled as `early
types' or E+S0s, and all galaxies with T=3,4,5 are labeled as spirals.

Cl~0024 and MS~0451 were respectively observed for 15ks and 80ks
with {\it GALEX} \citep{martin05} in 2004 October 
(GO-22; Cycle 1; PI Treu), reaching comparable depths in rest frame
FUV (observed NUV). Galaxy fluxes were measured within $6\arcsec$
circular apertures, centered on the optical position, on images
reduced and sky-subtracted using the standard automated 
{\it GALEX} reduction pipeline. The aperture size is chosen to be
comparable to the measured NUV FWHM ($5\farcs5$) to avoid difficulty
with source confusion, which can be a significant issue for 
{\it GALEX} imaging. This problem limits our ability to 
measure total fluxes in larger ($>6\arcsec$) apertures 
centered on the optical position, as contamination from nearby sources 
becomes much higher. As a result, we apply an aperture correction 
to bring our aperture fluxes into agreement with
SExtractor-derived total magnitudes ({\tt MAG\_AUTO}), and for comparison to
{\tt MAG\_AUTO} magnitudes in F814W \citep{bertin96}. We note that at
the redshifts of the clusters, most galaxies are effectively point
sources at {\it GALEX} resolution, so we do not expect that the
aperture corrections introduce significant errors into the photometry.

While observed F814W provides a good match to rest-frame $V$ for
both Cl~0024 and MS~0451, we also make use of supplemental
ground-based imaging to aid in determining k--corrections to transform
observed magnitudes to the nearest rest frame bands. 
Ground based data include panoramic $K_s$-band imaging of both 
clusters, along with $J$-band imaging of Cl~0024, obtained with the WIRC
camera on the Hale 200'' telescope at Palomar Observatory
\citep{wilson03}. These near-infrared data are supplemented 
with optical imaging in {\it BVRI} bands from either
the 3.6-m Canada-France-Hawaii Telescope (Cl~0024) or the Subaru 8-m
telescope (MS~0451), which have been described in \citet{moran07a}.

We use the {\sl kcorrect} software v.4\_1\_2 \citep{blanton03} to
estimate the necessary k--corrections; we make use of imaging in
all available bands from NUV to $K_s$, modeling each
galaxy's spectral energy distribution and deriving the best-fit
correction on a galaxy by galaxy basis. However, we find that the 
derived k--correction is mostly insensitive to the omission of one or
several bands, with scatter typically $\sim0.1$ magnitudes. 
We assume a Galactic extinction of E({\it B--V})=0.056 for Cl~0024 
and 0.033 for MS~0451 \citep{schlegel98}, and all absolute magnitudes will
be expressed on the AB system. 

We note also that, in all
cases where we quote rest-frame colors, we have applied a k--correction
equal to the {\it  median} k--correction for all galaxies of the same 
morphological type and in the same cluster, rather than the individually fit
k-corrections. We do this to avoid introducing any additional scatter
into the rest-frame colors (due to uncertainties in the k--corrections), 
above that present in the observed-frame colors. In general, the
difference between the two methods is $<0.1$ mag, and we do not expect
the method of k--correction to impact our analysis significantly.

\subsection{Spectroscopy}

\begin{deluxetable}{cccc}
\centering
\tablewidth{0pt}
\tablecaption{MS~0451-03 Redshift Catalog}
\tablehead{\colhead{$\alpha$} & \colhead{$\delta$} &
  \colhead{$z$} &
  \colhead{Source\tablenotemark{a}}  \\
\colhead{($^o$)} & \colhead{($^o$)} & \colhead{} & \colhead{}}

\startdata
73.321434 & -3.022260 & 0.667 & 1   \\
73.328568 & -3.033381 & 0.370 & 1   \\
73.325325 & -3.024078 & 0.300 & 1   \\
73.334717 & -3.043395 & 0.128 & 1   \\
73.331795 & -3.002323 & 0.539 & 1   \\
73.344551 & -3.028507 & 0.725 & 1   \\
73.346550 & -3.024269 & 0.371 & 1   \\
73.339874 & -3.009066 & 1.160 & 1   \\
73.337769 & -3.004604 & 0.388 & 1   \\
73.343781 & -3.007794 & 0.995 & 1   \\
... \\
\enddata
\tablenotetext{a}{Source codes: $1=$ Ellingson et al. (1998) , $2=$ Keck/DEIMOS}
\tablecomments{\label{0451cat} The complete version of this table is in the electronic edition of
the Journal. The printed edition contains only a sample.}
\end{deluxetable}

We have used the DEIMOS spectrograph on Keck II to obtain
deep spectra of $\sim 1500$ galaxies/cluster in the fields of
Cl~0024 and MS~0451. In both clusters, we observe with
$1\arcsec$ wide slits with lengths much larger than the galaxy size
(typically $8\arcsec-12\arcsec$). 
For each cluster, spectral setups were chosen to span rest frame wavelengths
from $\sim3500\mbox{\AA}$ to $\sim6700\mbox{\AA}$, covering optical
emission lines \oii, \oiii, H$\beta$, and, more rarely,
H$\alpha$.

Observations of Cl~0024 from 2001 October to
2003 October are described in Paper I and Paper III. Integrations
totaled 2.5hr per mask and targets were selected 
from the CFHT $I$--band mosaic, with priority
given to known cluster members with {\it HST} morphologies,
followed by galaxies in the {\it HST} survey without a known
redshift, to $I<22.5$. 

We undertook an initial redshift survey of MS~0451 during 2003 October,
observing 14 slitmasks for one hour each, with the 600 l~mm~$^{-1}$ grating
set to central wavelength 7500$\mbox{\AA}$. This allowed redshift
identification for 1300 objects, including 250 cluster members.
Targets were selected from the I-band Subaru image, 
as the ACS mosaic was not yet available. We first selected objects randomly
from the set with $I<21.5$, and then filled in slit mask gaps with
fainter objects ($I<23.0$).

We have additionally observed both clusters again in 2004 December
and 2005 October. For the 2004 December run, 
spectroscopic targets in Cl~0024 were selected in the same 
manner as previous runs. Integrations were 2.5hrs long, but used
the 600~l~mm$^{-1}$ grating set at $6200\mbox{\AA}$ central wavelength.

In 2004, most targeted objects in MS~0451 were selected for deeper followup
after already being identified as cluster members in our previous
year's redshift survey. However we excluded bright cluster members
that had already yielded spectra of sufficient signal to noise in
the 1-hr integrations of 2003. We observed 3 masks for 4 hrs each,
with gaps between high-priority targets filled as before with random
objects to $I<23.0$. Observations used a
600~l~mm$^{-1}$ grating set to a central wavelength of 6800$\mbox{\AA}$.

In 2005 October, objects in both clusters were observed with the same
spectral set up and integration times as in 2004, with six masks
observed per cluster. However, in 2005 our goal was to follow
up on sources detected in both {\it GALEX} imaging and {\it Spitzer
  Space Telescope} MIPS
imaging \citep{geach06}, and so we filled masks with these sources 
preferentially. While targets in these latest observations were not
selected randomly to a fixed magnitude limit, the total spectroscopic 
sample is still representative of the cluster as a whole, 
with the bulk of cluster members identified prior to 2005. 

\begin{deluxetable*}{ccccccccc}
 \centering
  \tablewidth{0pt}
  \tablecaption{\label{tab:data} Photometric and Spectroscopic Measurements for All
  Cluster Galaxies}
 \tablehead{ \colhead{RA}  &
  \colhead{DEC} & \colhead{z} & \colhead{Morph} & \colhead{F814W} &
  \colhead{$M_V$} & \colhead{\oii} & \colhead{$D_n(4000)$} & 
  \colhead{$H\delta_A$} \\
 \colhead{($^\circ$)}  &
  \colhead{($^\circ$)} & \colhead{ } & \colhead{ } & \colhead{mag} &
  \colhead{mag} & \colhead{$\mbox{\AA}$} & \colhead{ } & 
  \colhead{$\mbox{\AA}$}}
\startdata
 6.443974 & 17.141319 & 0.3792  &E/S0 & 21.9 & -18.8 & $ -25.6\pm  1.0$ & $1.10\pm0.02$ & $  1.6\pm  0.1$ \\
 6.462808 & 17.145109 & 0.3978  &  S0 & 19.1 & -21.7         & $   2.2\pm  0.0$ & $1.79\pm0.01$ & $$---$$ \\
 6.477189 & 17.274000 & 0.3967  &  S0 & 20.0 & -20.7 & $  -1.8\pm  0.2$ & $1.74\pm0.02$ & $  1.2\pm  0.1$ \\
 6.481423 & 17.247549 & 0.3969  &Sc+d & 20.7 & -20.1 & $ -30.8\pm  1.7$ & $1.20\pm0.02$ & $  5.6\pm  0.1$ \\
 6.499555 & 17.336269 & 0.3951  &  S0 & 19.9 & -20.8       & $   9.9\pm  4.1$ & $$---$$ & $ -3.8\pm  0.1$ \\
 6.508250 & 17.056641 & 0.3992  &Sa+b & 20.5 & -20.2 & $  -2.4\pm  0.4$ & $1.36\pm0.02$ & $  1.8\pm  0.2$ \\
 6.510695 & 17.325680 & 0.3931  &  S0 & 19.2 & -21.5 & $ -13.6\pm  0.5$ & $1.77\pm0.01$ & $ -1.8\pm  0.1$ \\
 6.511346 & 17.324520 & 0.3913  &  S0 & 19.7 & -21.0 & $  -9.4\pm  0.8$ & $1.34\pm0.02$ & $  4.6\pm  0.1$ \\
 6.514315 & 17.317190 & 0.3946  &   E & 18.8 & -21.9         & $   1.2\pm  0.4$ & $1.77\pm0.02$ & $$---$$ \\
 6.517361 & 17.105089 & 0.4059  &Sa+b & 18.8 & -21.9 & $   1.8\pm  0.5$ & $1.80\pm0.02$ & $ -2.1\pm  0.1$ \\
... \\

\enddata
\tablecomments{The complete version of this table is available in the
  electronic edition of the Journal. The printed edition contains only
  a sample.}
\end{deluxetable*}

DEIMOS data were reduced using the DEEP2 DEIMOS data reduction
pipeline \citep{davis03}, which produce sky-subtracted,
wavelength-calibrated one- and two-dimensional spectra.  Redshifts for all
galaxies were determined from the one-dimensional spectra and 
verified by eye. We further
correct spectra for the instrumental response curves before measuring
line strengths, but we do not perform flux calibration.

In total, we have obtained spectra of over 300
member galaxies per cluster, to $M_V=-18.0$, boosting the total known
cluster members to 504 in Cl~0024 and 319 in MS~0451. The
spectroscopic sample is $>65\%$ complete for objects with F814W$<21.1$
in Cl~0024 ($M_V=-19.6$ at the cluster redshift). In MS~0451, 
completeness to the same absolute magnitude limit is lower, $\sim30\%$, 
because of the deeper observations required (to F814W$=22.0$). In both
clusters, the spectroscopic sample remains representative of the
cluster population as a whole to $M_V=-19.6$, roughly 1.5 magnitudes
below $M_*$. However, in the range
$-19.6<M_V<-18.0$, we are biased toward detection of emission line
cluster galaxies over absorption line galaxies, and completeness is
lower in MS~0451 than Cl~0024. In the following, we will 
restrict our analysis to cluster members with $M_V<-19.6$, except
where specified otherwise. 

Our full redshift catalog for Cl~0024 was published in
Paper III, excluding the small number of additional redshifts
obtained in 2005. In Table~\ref{0451cat}, we publish a sample of 
the equivalent full catalog for MS~0451, containing redshifts for 1562
objects. The full catalog is available in the online edition.

As Cl~0024 and MS~0451
have now been observed in more detail than perhaps any others at
these redshifts, we aim to make as much of our data as practical
available to other investigators through our web site.
We will therefore publish updated versions of both the Cl~0024 and MS~0451
catalogs to our web
site\footnote{http://www.astro.caltech.edu/$\sim$smm/clusters/}, including
positions matched with photometric measurements, {\it HST}
morphologies, and redshifts.

\subsection{Spectral Line Measurement and Velocity Dispersion}

We measure spectral line indices for several key
emission and absorption lines following the Lick system
\citep{worthey94}. We focus in this paper on indicators of 
recent or ongoing star formation, including \oii \ 3727 and the 
Balmer lines H$\delta_A$ and H$\beta$, which have been described 
in Paper III. We also measure the $D_n(4000)$ index, which indicates 
the strength of the 4000$\mbox{\AA}$ break; we adopt the
definition from \citet{balogh99}.

In order to make maximal use of our spectra for 
the investigation of stellar populations and star formation rates in
cluster transition galaxies, we took great care to measure the
equivalent widths (EWs) of key spectral lines in an optimal way.
We developed a code in IDL which measures EWs via an inverse-variance
weighted integration of the flux across the line. 
The inverse variance ($1/\sigma^2$)
for each pixel in a spectrum is automatically generated as
output from the DEIMOS reduction pipeline. Weighting by inverse
variance minimizes the effects of poorly subtracted sky lines or other
low S/N regions of the spectrum in the calculation of EWs, and allows
for accurate uncertainty estimates on the indices. The code also
robustly measures {\it lower limits} on emission line EWs in spectra
where emission lines are bright but the underlying stellar continuum is
undetected. As in earlier papers, we adopt the convention that
emission lines have negative equivalent widths and absorption lines
positive. Hereafter, we refer to the equivalent widths in the \oii \ 
line as simply ``\oii'', and we refer to H$\delta$ equivalent widths
by the index name, ``H$\delta_A$''.

In total, we measure EWs for samples of 116 and 124
spiral galaxies in MS~0451 and Cl~0024, respectively, as well as
samples of 130 and 109 E+S0s in MS~0451 and Cl~0024. Photometric
characteristics and key spectral line measurements for each of these
galaxies are listed in Table~\ref{tab:data}. 

While our spectroscopic
sample of cluster members contains no broad-lined AGN, 
narrow-lined AGN are harder to 
identify due to the lack of H$\alpha$ in most spectra, complicating
identification based on spectral line ratios. In \S4, we therefore
must consider the possibility of AGN contamination through other means.

In determining the Fundamental Plane of early type galaxies (\S6), we
require measurement of the line of sight velocity dispersion from the
spectra of E+S0 cluster members. Despite a variety of spectral setups,
as detailed above, all of our DEIMOS observations yield spectra with 
a velocity resolution of $\sim30-50$~km~s$^{-1}$.
As described in  Paper III for Cl~0024, we measure velocity
dispersions for MS~0451 early types using the Gauss-Hermite
pixel-fitting code of \citet{pixfit}, limiting our sample to those
observed galaxies with spectral signal to noise (S/N) $>7$ 
($\mbox{\AA}^{-1}$), with a median S/N$=13$($\mbox{\AA}^{-1}$). 
Extensive testing
of this code discussed in Paper III and \citet{tt05b} imply errors of
$<10\%$ in the derived velocity dispersions for galaxies above this
S/N cutoff. In total, 60 E+S0s in MS~0451 and 71 in Cl~0024 have 
high S/N spectra suitable for measurement of the velocity dispersion.

\section{Passive Spirals}

The discovery of passive spirals in clusters at intermediate redshift
\citep{couch98,d99, poggianti99} has led to their recognition
as promising candidates for the role of `transition object' during
the theorized conversion of cluster spirals into S0s between $z\sim0.5$ 
and today. Such a conversion, though controversial \citep[e.g.,][]{burstein05}, 
has been expected because of the striking 
contrast between the large number of spirals observed in 
clusters at $z\sim0.5$ and the correspondingly large population 
of S0s found in clusters locally. 
In our initial analysis of these passive spirals in
Cl~0024 \citep{moran06}, we found evidence in their stellar
populations for a slow decline in star formation, confirming that
passive spirals--at least in Cl~0024--exhibit the prerequisite 
cessation of star
formation  needed for any transformation into S0s. 
Perhaps more importantly, we found that the {\it abundance} of passive 
spirals in Cl~0024 is quite high, over 25$\%$ of the total spiral 
population--enough, perhaps, to account for the {\it entire} buildup
of S0s in clusters \citep[but see][]{kodama01}.

 However, several questions remain about the nature
of passive spirals and their purported S0 end states. Among them:
Apart from their lack of emission lines, how do the passive spirals 
as a class differ from the star forming spirals, and are their 
properties uniform between clusters? 
If they are truly destined to transform into S0s, can we identify 
directly these new S0s, perhaps just after their new morphologies become
firmly in place?
Finally, in what environments do the passive spirals reside, and with what 
implications for the physical mechanisms driving their creation? 
In this section, we attempt to answer the first of these questions,
specifying more fully the properties and recent star formation
histories of the passive spirals. In \S5 and \S6, we will then 
consider the next two questions, in our attempt to piece together the
full evolutionary history of Cl~0024 and MS~0451 galaxies.

We define as `passive' any spiral with an \oii \ equivalent width
\oii $>-5\mbox{\AA}$. Though this definition necessarily allows some
contamination from star-forming galaxies that are either moderately dusty or
forming stars at a low rate, we will demonstrate below via 
coadded spectra that all emission lines are weak or absent in the
passive spirals. We note that we would classify as `passive' 
many of the spiral galaxies in local clusters that are observed to be 
HI-deficient but retain a low level of star formation \citep{gavazzi06}. 
By examining multi-color (NUV and F814W) imaging of
the cluster by eye, we remove from the sample all galaxies where the
UV flux is likely to be significantly contaminated by neighboring objects or
image artifacts.

\begin{deluxetable}{lcccc}
  \centering
  \tablewidth{0pt}
  \tablecaption{\label{tab:uv}Passive Spirals and UV emission.}
 \tablehead{ \multicolumn{3}{l}{Fraction of UV-detected:}  &
  \colhead{} & \colhead{$N_{passive}/$}\\
\colhead{ } & \colhead{Active Spirals} & \colhead{Passive Spirals} &
\colhead{E+S0s} & \colhead{$N_{spirals}$ } }
  \startdata
Cl~0024 & $86\% \pm 6\%$ & $65\% \pm 14\%$ & $25\% \pm 7\%$ & $28\%
\pm 7\%$ \\
MS~0451 & $88\% \pm 5\%$ & $37\% \pm 15\%$ & $8\%\pm 3\%$ & $26\% \pm 6\%$ \\

\enddata
\tablecomments{Above calculations are restricted to $M_V<-19.6$, with
  passive and active spirals defined as described in the text.}
\end{deluxetable}

We also restrict our passive spiral sample 
to those objects with H$\delta_A<5.0\mbox{\AA}$; we assign otherwise
passive spirals with H$\delta_A$ stronger than this to the 
`post-starburst' category. We note that this H$\delta$ limit 
differs from the definition adopted by some other authors
\citep[e.g.,][]{poggianti99}, who label all passive galaxies with
  H$\delta>3\mbox{\AA}$ as post-starburst. 
However, the difference is primarily due to different measurement
methods for H$\delta$: we find that 
a large number of star-forming and passive spirals 
exhibit $3\mbox{\AA}<$H$\delta_A<5\mbox{\AA}$ in our sample, 
which is within the range where \citet{kauffmann03}
report that star formation histories are consistent with continuous 
star formation.
Therefore, spiral galaxies in our sample with \oii$>-5\mbox{\AA}$ 
and H$\delta_A$ between $3\mbox{\AA}$ and 5$\mbox{\AA}$ are perhaps more
properly  called `post-{\it starforming}', as there is no need to invoke a
starburst to explain their H$\delta_A$ values. We therefore group them
with the remainder of the passive spirals.

\begin{figure}[t]
\centering
\includegraphics[width=3.25in]{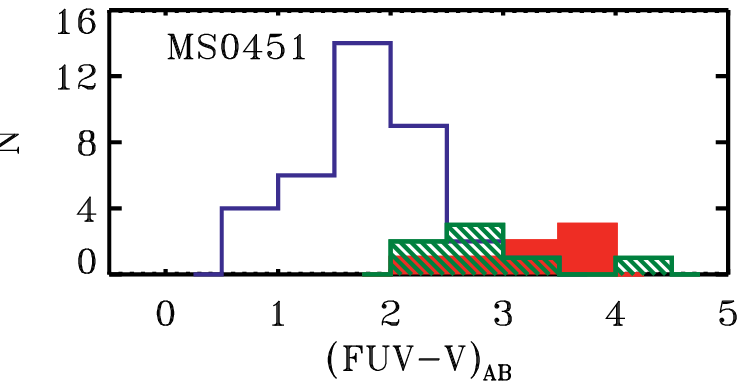} \\
\includegraphics[width=3.25in]{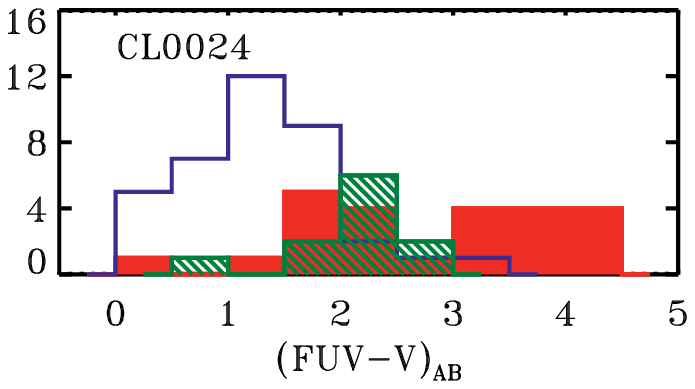}
\caption{\label{uvcolor}Distributions of FUV{\it~--~V} colors in 
  MS~0451 (top) and Cl~0024 (bottom). Open blue histograms 
  indicate the colors of normal star-forming spirals. 
  Green hatched histograms display the colors for those passive
  spirals that are detected in the UV, and solid red histograms
  likewise indicate the colors of E+S0s that have been detected in the
  UV. No correction for dust extinction is applied. 
  UV-detected passive spirals in both clusters have intermediate
  colors between those of star forming spirals and those few early
  types detected in the UV.}
\end{figure}

\begin{figure*}[th]
\centering
\includegraphics[width=2\columnwidth]{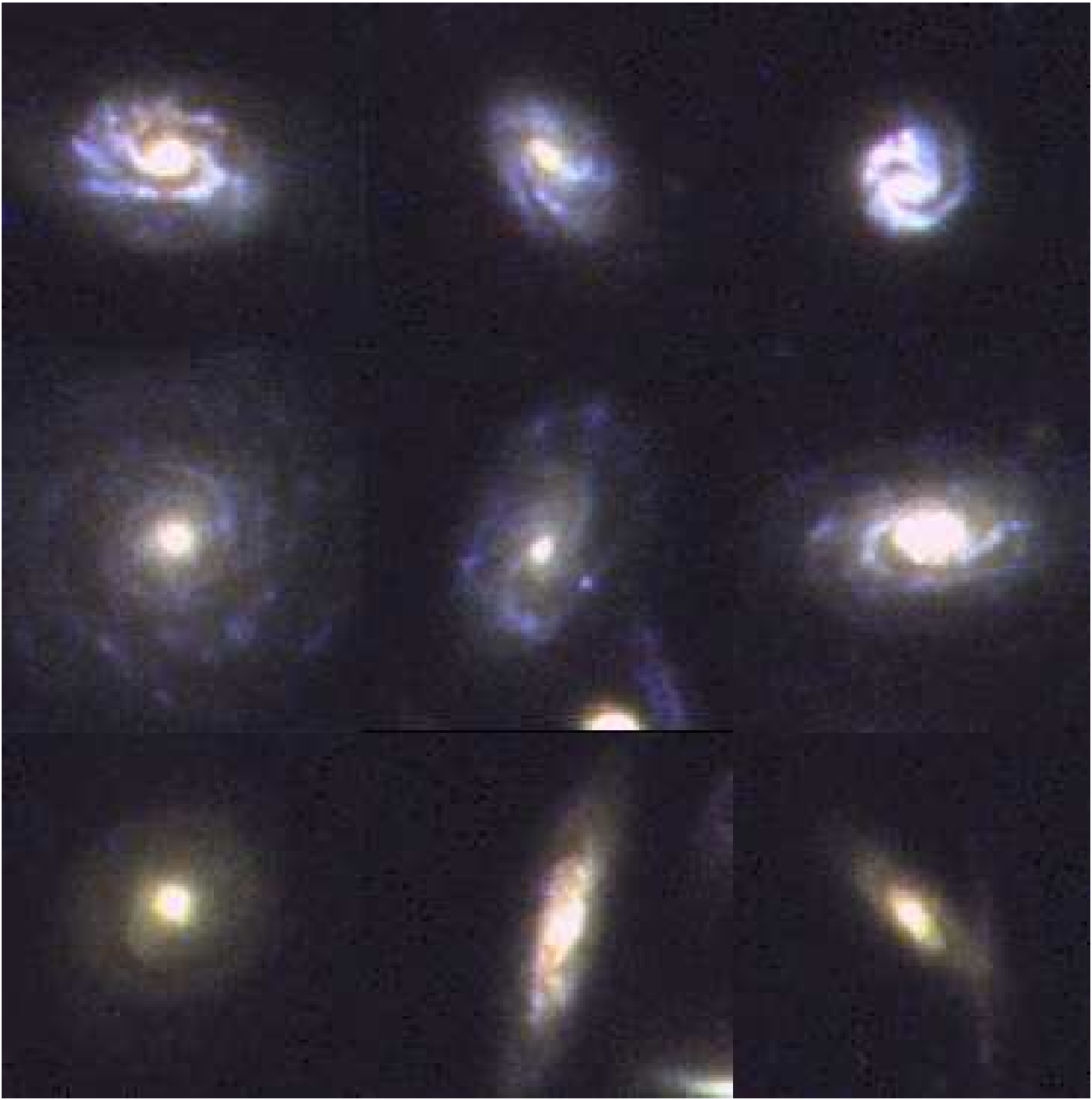}
\caption{\label{montage} Montage of active and passive spirals from deep multi-color
  ACS imaging of the core of Cl~0024.
  The top row displays three
  typical spirals whose spectra exhibit emission lines. The next two
  rows are composed of passive spirals, which qualitatively exhibit
  two forms: those with blue disks, but with possibly lower surface
  brightness than star-forming spirals (middle row), and those with
  distinctly red disks (bottom row). All galaxy images are $4\arcsec
  \times 4\arcsec$ and were extracted
  from the same multi-color image with identical image scaling and
  color balance, with F850LP in red, F775W in green, and F555W in blue.}
\end{figure*}

After culling contaminated objects and dividing our sample 
according to their \oii \ and H$\delta_A$ strengths, we find that $87\%\pm5\%$ of
MS~0451 spirals with \oii \ emission (brighter than $M_V=-19.6$) 
are detected in our {\it GALEX} 
imaging, similar to the detection rate in Cl~0024. Conversely, only
$37\%\pm15\%$ of spectroscopically passive spirals are similarly
detected (dropping to $32\%\pm13\%$ if we were to include the
post-starburst spirals), significantly less than 
the UV-detected fraction in Cl~0024 (see Table~\ref{tab:uv}). 
As the rest-frame luminosity limits for detection of FUV are
virtually the same between the two {\it GALEX} exposures ($M_{FUV}<-17.6$ in
MS~0451 and $M_{FUV}<-17.9$ in Cl~0024), this represents an important
difference in UV luminosity between the two clusters' passive spiral 
populations. 

In Figure~\ref{uvcolor}, we display the distributions of rest-frame
FUV{\it~--~V} colors for star-forming spirals, passive spirals, and early
types in each cluster. It is clear that, despite the overall lower
fraction of UV-detected passive spirals in MS~0451, those that are
detected exhibit the same intermediate colors as in Cl~0024.
To further investigate the colors of passive spirals, and to include
those that were not UV-detected, we also utilize our ground-based {\it
R} and {\it I}-band imaging to  measure rest-frame {\it B--V}, 
for passive and normal spirals. 
We observe a trend similar to that seen in the UV--optical colors: 
the median passive spiral is $\sim0.1$ magnitudes redder than the median
star-forming spiral in Cl~0024, and 0.2 magnitudes redder in MS~0451. 
KS tests indicate that passive spirals have a color distribution 
significantly different from that of normal spirals, in both
clusters. 

In fact, we can verify visually that passive spirals appear
different from normal star forming spirals.
In Figure~\ref{montage}, we display color postage stamp images of nine
spirals in Cl~0024, derived from deep multicolor ACS imaging of the
center of Cl~0024 which we retrieved from the {\it HST} archive
(GTO Proposal 10325, PI:Ford). 
In the top row, we display three normal star-forming cluster
spirals. The next two rows show cluster passive spirals, which appear to fall
into two basic types. Shown in the middle row, some passive spirals
appear to retain blue disks, but at surface brightnesses lower than
that of star-forming spirals. While some star formation may be
occurring in these galaxies, we note that our spectroscopic slits are 
large enough that
we do not believe we are simply `missing' regions of star formation in
our observations. 
In the bottom row, we see three examples
of passive spirals with distinctly red disks, yet with spiral arms and
dust lanes still present. These appear similar to the $z=0$ `anemic
spirals' first identified by \citet{vandenbergh76}; indeed, inspection
of the integrated spectrum of a prototypical `anemic spiral',
NGC~4569 \citep{gavazzi04}, indicates that it would likely
satisfy our definition of `passive spiral'. 
It seems, then, that passive spirals
could genuinely be disk-dominated systems where star formation is 
on the decline or has halted.

In counterpoint to their lower incidence of UV emission, the overall
frequency of passive spirals in MS~0451 is quite similar to that found
in Cl~0024 (Table~\ref{tab:uv}). In both clusters, more than a quarter of all spirals are
passive--important confirmation that passive spirals are a significant
component of both clusters despite the large difference between the
clusters' global properties. 

Yet their presence in two quite distinct clusters
leads us to question whether the passive spirals are a cluster-related
phenomenon at all, or if they could instead represent some fraction
of all spirals that have internally exhausted their star formation.
If they are indeed generated in the cluster environment, then 
we would not expect to find passive spirals in the field at these 
same intermediate redshifts. We have therefore examined a sample 
of 105 field spirals in the redshift range $0.3<z<0.65$, identified 
in the course of our spectroscopic campaign. Out of 62
galaxies where \oii \ falls within our wavelength coverage, we
measure a passive spiral fraction of only $6\%\pm3\%$. This low
incidence in the field confirms that the generation of passive 
spirals is a cluster-related phenomenon at these redshifts
\citep{poggianti99}. The
$6\%\pm3\%$ of passive spirals found in the field may indicate that
some passive spirals can be formed in groups.

Despite their overall similar abundance in both clusters, the 
weaker UV emission in MS~0451 passive spirals presents a 
puzzle that leads us to again consider the different assembly 
states and ICM properties of the two clusters.
To more precisely quantify the nature of the passive spirals 
in each cluster, we introduce as a key diagnostic plot 
the FUV{\it~--~V} vs $D_n(4000)$ diagram. 
$D_n(4000)$ is sensitive to stellar
populations with ages of $\sim2$~Gyr, with a dependence on
metallicity that only becomes apparent for old stellar populations 
\citep{poggianti97}, while the FUV{\it~--~V} color
is sensitive to star formation on a much shorter timescale
(10$^7$--10$^8$ yr). As we will see, comparing $D_n(4000)$ to 
FUV{\it~--~V} color provides a valuable tool to discriminate 
between different star formation histories for passive and
star-forming spirals.

\begin{figure}
\centering
\includegraphics[width=\columnwidth]{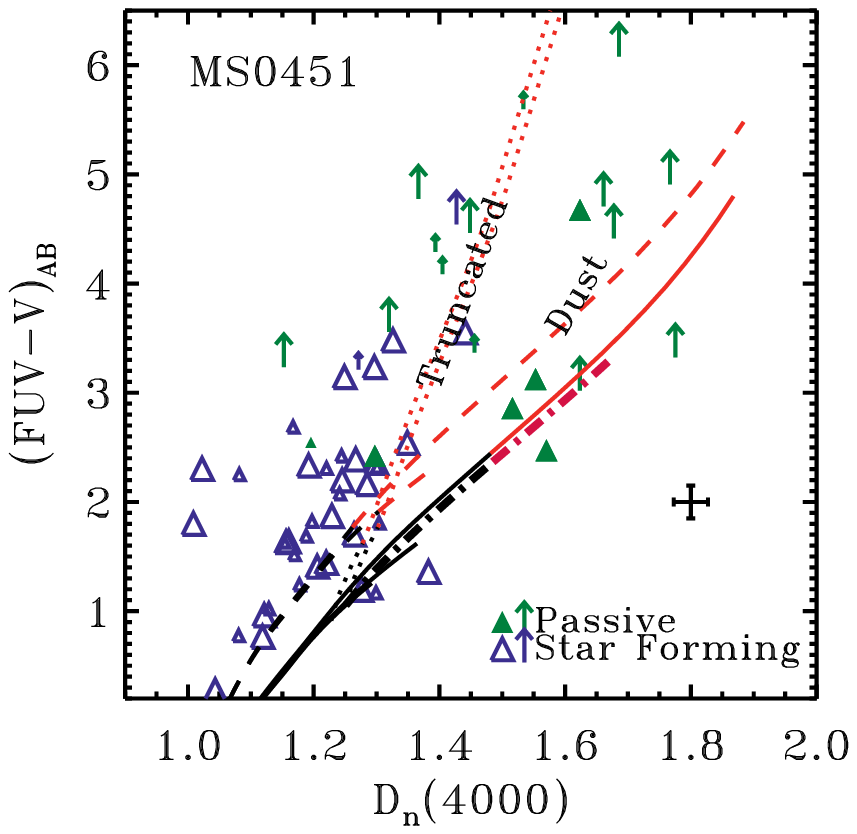}
\includegraphics[width=\columnwidth]{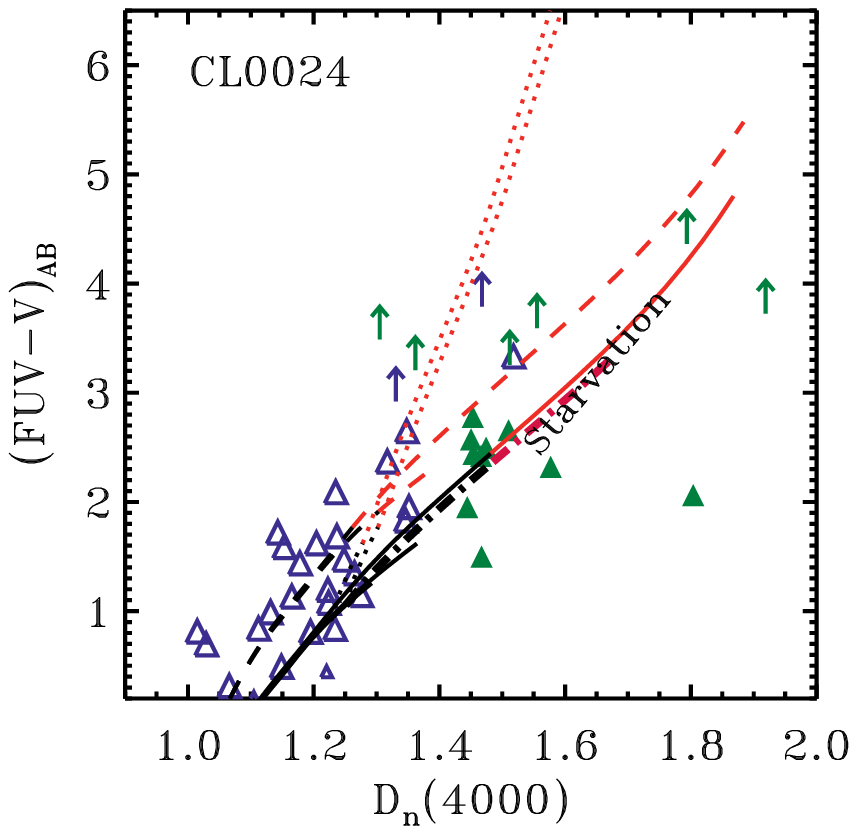}
\caption{\label{hd_uvnew} 
FUV{\it~--~V} colors versus $D_n(4000)$ strength for spirals in MS~0451
(top) and Cl~0024 (bottom). Star-forming and passive spirals are 
indicated by open blue triangles and solid green triangles,
respectively. Arrows denote
3-$\sigma$ lower limits for FUV{\it~--~V} colors for those galaxies
not detected in the UV, colored blue for star-forming spirals and
green for passive spirals. Small triangles or lower limit arrows
indicate spirals with H$\delta_A>5\mbox{\AA}$, similarly
color coded. We plot only galaxies with especially small uncertainty
in $D_n(4000)$, $<0.07$.
Solid, dashed, dotted, and dash-dotted lines respectively indicate 
model tracks for exponentially declining, dust-enshrouded, truncated,
or `starved' star formation, as described in the text. Tracks are
colored red for regimes where \oii$>-5\mbox{\AA}$. Most UV-detected
passive spirals are consistent with a starvation model, while most
undetected passive spirals are better fit by a rapid truncation of star formation.}
\end{figure}

In Figure~\ref{hd_uvnew}, we plot $D_n(4000)$ versus rest frame 
FUV{\it~--~V} color for MS~0451 (top) and Cl~0024 (bottom). 
It is clear from the diagram that most star-forming spirals in 
both clusters exhibit 
blue FUV{\it~--~V} color (median of $1.6\pm0.1$) combined with weak
$D_n(4000)$ strength (median $1.22\pm0.02$).
Such values indicate young stellar populations, and 
are expected for galaxies with ongoing star formation.

Turning to examine the locations of passive spirals, we note that
passive spirals in MS~0451 largely occupy a different region
of the plot than Cl~0024 passive spirals. Cl~0024 passive
spirals with UV detections--the majority--cluster quite tightly 
on the plot at moderate FUV{\it~--~V} colors (median $2.5\pm0.1$) 
and moderately strong $D_n(4000)$ (median $1.47\pm0.02$). 
In contrast, as we have already noted, many of the
MS~0451 passive spirals have only upper limits on their FUV
emission, some at quite red FUV{\it~--~V} colors. Across both
clusters, these UV-undetected objects seem also to exhibit a wider 
spread in $D_n(4000)$ than the UV-detected passive spirals (rms of $0.23\pm0.06$ c.f. $0.08\pm0.02$). The
dominance of UV-detected passive spirals in Cl~0024, compared to the
UV-undetected type that are predominant in MS~0451, begs the question
of whether passive spirals are created in different ways in each of
the two clusters. As we will show, the two types of passive
spirals--UV detected and undetected--may have fundamentally 
different star formation histories.

To help decipher the star formation histories of passive
spirals in Cl~0024 and MS~0451, we overlay in Figure~\ref{hd_uvnew}
evolutionary tracks from the population synthesis code of
\citet{bc03}, adopting fixed solar metallicity for all models. 
Using these, we can test several classes of models that
could explain the origin of passive spirals. Specifically, can passive
spirals be created by simply adding dust to a star-forming spiral? Can
they be created by suddenly switching off star-formation in a galaxy,
which we will call `truncation' models? Or, finally, can we create the
passive spirals by switching a star-forming spiral into a phase where
its star formation rate declines more rapidly over a timescale of
0.5--2~Gyr (a `starvation' model)?

In Figure~\ref{hd_uvnew}, solid lines indicate the track followed in 
FUV{\it~--~V} vs. $D_n(4000)$ space by several model stellar
populations
with exponentially-declining SFRs, with characteristic timescales
$\tau\sim3$--7~Gyr. Such models have been found to reproduce the 
optical spectra of Sa-Sc--type star forming galaxies \citep{poggianti96}.  
Dashed lines add dust extinction of $A_V=0.6$ to these same models, 
corresponding to the mean difference in FUV{\it~--~V} between 
the (UV-detected) passive and active spirals. 
Active spirals largely occupy the 
region in between tracks with zero and moderate extinction.
Tracks are in red for regimes where \oii$ > -5 \mbox{\AA}$,
estimated from the models in the same way as in \citet{moran06}.

Only models with a relatively fast 
exponential decline in SFR ($\tau\le1$~Gyr, upper solid and dashed lines in 
Figure~\ref{hd_uvnew}) can match the peculiar combination of
intermediate FUV{\it~--~V} and strong $D_n(4000)$ exhibited by 
the bulk of the UV-detected passive spirals. Any star forming spirals that
transition into this sort of rapid decay of star formation,
eventually entering a spectroscopically passive phase, 
will reproduce the positions of the UV-detected passive spirals in
the diagram (thick dash-dotted `Starvation' line in Figure~\ref{hd_uvnew}). 

Conversely, passive spirals with lower limits on their color, which 
dominate the population in MS~0451, seem to be most consistent with
models where star formation is rapidly truncated at an age $<7$~Gyr,
indicated by the dotted lines in Figure~\ref{hd_uvnew}. Stacking
together the UV images for undetected passive spirals, we still
detect no significant UV emission. Though confusion noise becomes
significant in the stacked image, the nondetection implies that the
median FUV{\it~--~V} is at least $\sim1$ magnitude fainter than the
upper limits indicated in Figure~\ref{hd_uvnew}. 

The positions of these UV-undetected spirals on the plot could 
also be explained by strong levels of dust obscuration.
However, 24$\mu$m imaging of both clusters with MIPS on the 
{\it Spitzer Space Telescope} indicates that there is a deficit of obscured 
dusty starbursts in MS~0451, in comparison to Cl~0024
\citep{geach06}. If the UV-undetected passive spirals were simply
dusty, we would expect to see the {\it opposite} trend in the MIPS
observations, since there are so many of these objects in MS~0451. 
We therefore believe that the rapid truncation of star formation 
is the most likely explanation for the UV-undetected passive spirals.

The population of spirals in MS~0451
includes a number of post-starburst galaxies that also seem to 
be consistent with rapidly-truncated star formation; in fact, rapid
truncation of a starburst is thought to be the primary way that these
galaxies achieve such high H$\delta_A$ values
\citep{poggianti99}. Indeed, the post-starburst spirals may be
closely related to the UV-undetected passive spirals in MS~0451. 
Several apparently star-forming galaxies also
reside along the `truncated' track. Together, these could represent 
a continuum of galaxies in various stages of
having their star formation halted. However, the supposedly rapid
timescale for the cessation of star formation begs the question of why
we would see any star-forming spirals in the region of the plot
where models indicate that they should already be passive. Some mix of
models, with increased dust obscuration combined with the 
truncation of star formation, could provide an explanation.

While we have found passive spirals to be abundant in both Cl~0024 and
MS~0451, it appears that they are largely formed through different mechanisms
in each cluster. The more rapid cessation of star formation required
to explain the MS~0451
passive spirals is likely due to some physical mechanism that exerts a
stronger force on galaxies in MS~0451 than in Cl~0024. This once again
brings to mind the hot, dense ICM of MS~0451, which at least has the 
potential to apply a much stronger force (ram-pressure stripping) 
on infalling galaxies than in Cl~0024. In the next section, we will 
examine the stellar populations of early type galaxies across the two
clusters, in an attempt to identify the expected end products of these
rapidly and slowly quenched passive spirals: S0s with signatures of
recent star formation on varying timescales.

\section{THE STAR FORMATION HISTORIES OF E+S0 GALAXIES}

While careful analysis of local S0 galaxies has revealed some signs
that their stellar populations are younger than those of ellipticals \citep[e.g.,][]{poggianti01},
such signatures have so far been elusive at intermediate redshift, despite
expectations that S0s with newly halted star formation would be
abundant. Now, however, under the hypothesis that passive spirals in
MS~0451 and Cl~0024 must be disappearing from our spiral sample as they
fade in UV and increase in $D_n(4000)$, we turn again to a study of
the stellar populations of early type galaxies.

\begin{figure*}
\centering
\includegraphics[width=6.5in]{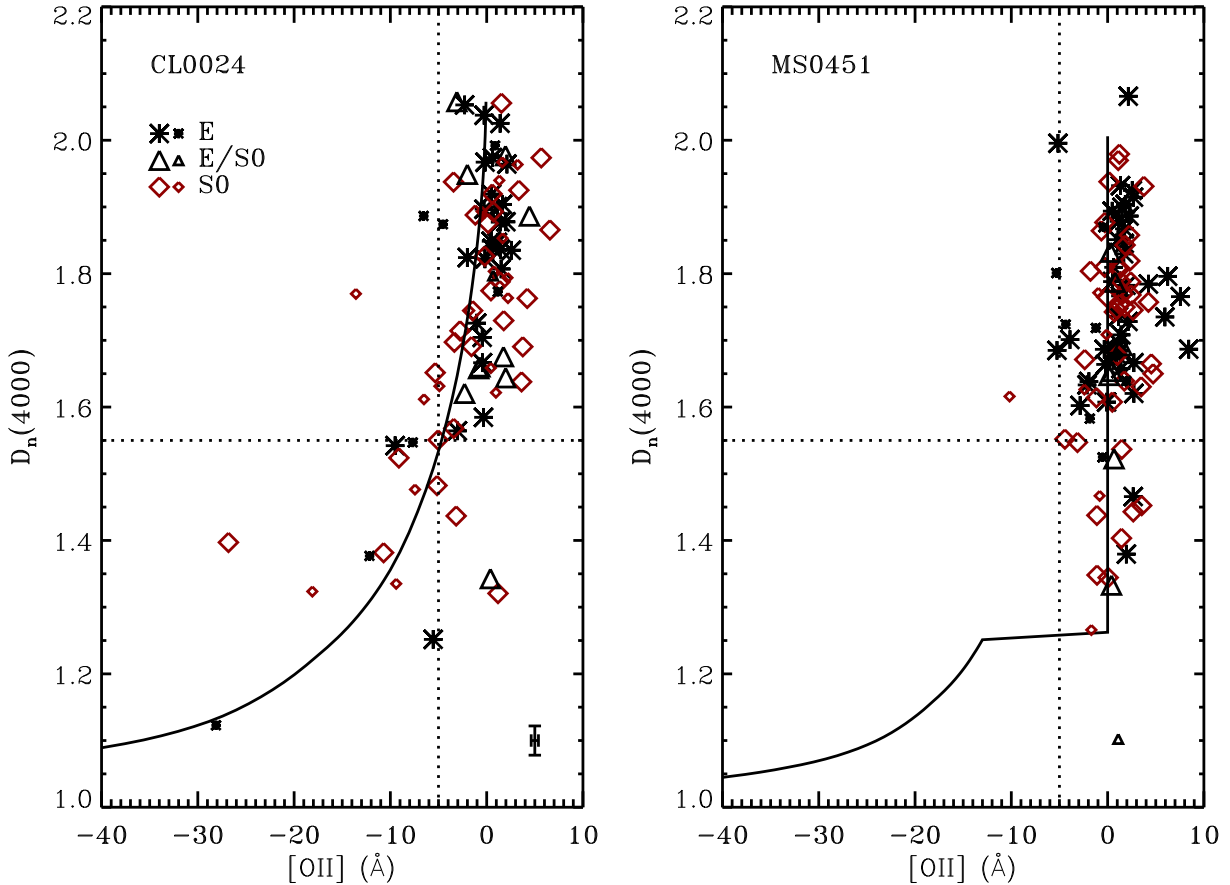}
\caption{\label{dn4000_oii} $D_n(4000)$ Balmer break strengths versus
  \oii \ 
  equivalent widths for cluster early types in Cl~0024 (left) and
  MS~0451 (right). Symbol shapes and colors denote galaxy morphologies
  as indicated on the figure, with median error bar indicated at lower
  right (left panel). Small symbols indicate galaxies that are
  located outside the cluster virial radius, and large symbols mark
  galaxies within $R_V$. Vertical and horizontal dotted lines
  respectively indicate \oii$=-5\mbox{\AA}$ and $D_n(4000)=1.55$
  (See text). 
  Solid line in the left panel is the best-fitting model track,
  a starvation ($\tau=1$~Gyr) model with dust, while in the right-hand
  panel the best fit is to a truncation model (at age 5~Gyr). The
  best-fit model tracks for each cluster are similar to those seen in
  Figure~\ref{hd_uvnew}. }
\end{figure*}

The different characteristics of passive spirals in Cl~0024 and
MS~0451 help to provide a key for uniquely linking the passive spirals
to their potential S0 descendants. Because the properties of 
passive spirals largely differ between the two clusters, so too 
will any S0s that have just recently changed their morphology.

With this in mind, we wish to examine recent star formation as
well as any ongoing star formation in the cluster E+S0s. In
Figure~\ref{dn4000_oii}, we plot $D_n(4000)$ strength versus \oii \ 
equivalent width
for all early types in Cl~0024 (left) and MS~0451 (right), with
symbols coded to indicate their morphologies (as shown in the
legend).

Most bright E+S0s in both clusters exhibit the
signatures of old stellar populations: weak or absent \oii \ and a
strong Balmer break ($D_n(4000)\gtrsim 1.6$). In Cl~0024, however,
we observe a striking tail of galaxies extending from the locus of old
E+S0s toward weaker $D_n(4000)$, coupled with significant \oii \ 
emission. MS~0451 exhibits a similar tail of early type galaxies
toward weak $D_n(4000)$ strength, but {\it without} any associated
\oii \ emission.

If we apply a cut at $D_n(4000)=1.55$, indicated by the horizontal
dotted lines in the figure, we find that 10--15$\%$ of the total E+S0 
population in each cluster has $D_n(4000)$
below 1.55. While somewhat arbitrary, the chosen $D_n(4000)$
dividing line lies well below the expected value for an early type 
galaxy that has been passively evolving since formation at $z\ge 1$. In fact, in
the case of a prototypical elliptical whose stars were created in a
short burst, $D_n(4000)$ should fade from $1.25$ to $1.7$ in
less than $\sim2$~Gyr, according to \citet{bc03} models--less than the
time from $z=1$ to $z\sim0.5$.

There are essentially two possible explanations for finding E+S0 galaxies
with such low $D_n(4000)$: either they have undergone recent star formation, 
or they harbor an AGN that contributes to the spectrum. If they have formed
stars recently or are continuing to form stars, then they must
have either formed a significant population of stars
at $z<1$ while already displaying early type morphology, or else
transformed morphology from a star-forming spiral to an early type
(either singularly or through a merger).
Under each of these scenarios, there must be some way to account for
the presence of emission lines in the Cl~0024 objects but a lack
thereof in MS~0451.

A third possibility--that we have simply misclassified some spiral
galaxies as S0s--is easy to discount based on this last
requirement. If the low $D_n(4000)$ S0s are truly just spirals, then we
would not expect in MS~0451 to {\it exclusively} miscast as S0 just
those spirals with no \oii \ emission, while at the same time
misclassifying spirals with a range of \oii \ in Cl~0024. 
On the other hand, if we are biased towards
mixing up the morphological classifications of passive spirals and
S0s, this only strengthens the notion that one type may be
transforming into the other.

To better assess the star formation histories of these peculiar objects, 
we overplot in Figure~\ref{dn4000_oii} two characteristic model
tracks equivalent to those plotted for the passive spirals in
Figure~\ref{hd_uvnew}. In Cl~0024, we plot a starvation-like track
($\tau=1$~Gyr with internal $A_V=0.6$), while in MS~0451, we plot the
track of a galaxy with star formation truncated rapidly at an age of
5~Gyr. In each case, the model track reproduces the positions of the
galaxies very well. We note that most physically plausible 
truncated-spiral tracks are inconsistent with the distribution of 
points in Cl~0024, as no such models reach $D_n(4000)\gtrsim1.4$
before truncation, yet we still see galaxies with \oii \ emission in
this range.  Remarkably, then, a single starvation track
reproduces the positions of both the low-$D_n(4000)$ early types 
and most passive spirals (Figure~\ref{hd_uvnew}) in Cl~0024, while
a truncated star formation track similarly matches the star formation
histories of both classes in MS~0451. These similarities in star formation
history suggest an evolutionary link between the passive spirals
and the low-$D_n(4000)$ early types in each cluster.

\begin{figure}
\centering
\includegraphics[width=3.25in]{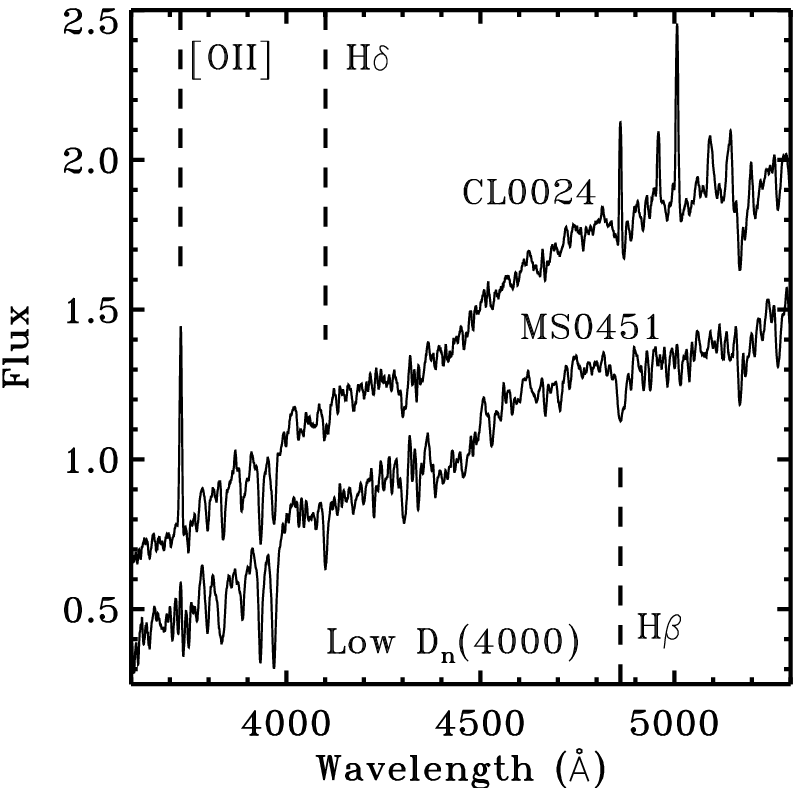}
\\
\includegraphics[width=3.25in]{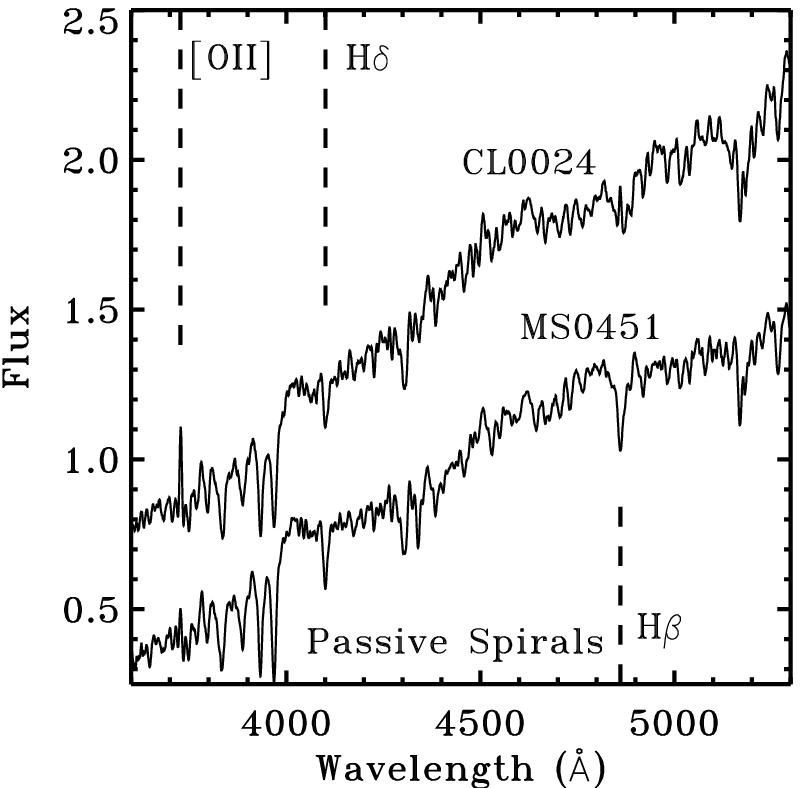}
\caption{\label{coadded} Top: Normalized, coadded spectra of 
  E+S0s in Cl~0024 and MS~0451 with $D_n(4000)<1.55$, 
  including only spectra with $S/N>5$ ($\mbox{\AA}^{-1}$). Cl~0024
  spectrum is shifted upward in flux (arbitrary units) for display
  purposes. The locations of several key spectral lines are marked. Bottom: 
  Normalized, coadded spectra of passive spirals in Cl~0024 and
  MS~0451, also restricted to spectra with $S/N>5.0$. In each cluster,
  the \oii, H$\delta_A$, and H$\beta$ strengths of passive spirals closely
  resemble those of the low-$D_n(4000)$ early types in the same cluster.
}
\end{figure}

We can further evaluate the likelihood of a connection between the
passive spirals and the low-$D_n(4000)$ early types 
through a consideration, for each
cluster, of the coadded spectra of both classes.
In Figure~\ref{coadded}, we plot in the top panel the coadded spectra
of all E+S0s in the clusters with $D_n(4000)<1.55$, and in the bottom
panel, the coadded spectra of passive spirals in each cluster. In each
case, we only include galaxies with overall spectral $S/N >5.0$
($\mbox{\AA}^{-1}$) in the summation. The locations of several key
spectral lines are marked.

Considering the coadded early-types first, we can immediately
cast doubt on the idea that AGN are responsible
for the low $D_n(4000)$. While a substantial contribution from a flat AGN
continuum can serve to weaken the Balmer break in a galaxy, the same
effect should also dilute the observed depth of stellar absorption
lines, such as H$\delta_A$. Yet it is clear from the coadded spectra
that the low-$D_n(4000)$ early types in both clusters exhibit moderately strong
H$\delta_A$: $2.2\pm0.2\mbox{\AA}$ and $1.2\pm0.2\mbox{\AA}$ for MS~0451
and Cl~0024, respectively. As both values are much higher than the
median H$\delta_A=-1.0\pm0.2$ of cluster ellipticals, their Balmer
strengths are consistent with the interpretation that they
contain young stars, and {\it inconsistent} with a significant AGN
(dust obscured or otherwise) contributing to the galaxy's spectrum.

Though the spectra make clear that these `young S0s' (as we will call
them) truly contain a
population of young stars, is it possible that
these galaxies have recently undergone a `rejuvenation' of star
formation through some interaction? 
Serious doubt is cast on this hypothesis by the
lack of MS~0451 E+S0s currently containing emission lines in their
spectra. After a moderate starburst on top of an established stellar
population, \citet{bc03} models predict a return to a passive spectrum
with strong $D_n(4000)$ in as little as 100~Myr after the burst. 
To have so many S0s experience a rejuvenation episode so recently, 
yet observe none of
them currently in a starburst phase, is implausible.

We are left, then, with the possibility that these young S0s truly
represent the end states of the passive spirals. 
The contrast between young S0s in Cl~0024, whose \oii \ strengths are
consistent with a gradual decline in star formation, and those in
MS~0451, where the lack of \oii \ indicates recent truncation, 
evokes the similar dichotomy between passive spirals with a slow
truncation of star formation (mostly in Cl~0024) and those exhibiting
a more rapid truncation (mostly in MS~0451).
Comparing the coadded
spectra of the two classes in more detail, we find further evidence to support
an evolutionary link between passive spirals and young S0s.

First, the passive spiral spectrum for
Cl~0024 exhibits a weak \oii \ emission line of $-3.1\pm0.5\mbox{\AA}$,
despite the population having been selected for their low
\oii.  The presence of some \oii \ emission in both the
passive spirals and the young S0s (\oii$=-10.1\pm0.5\mbox{\AA}$)
is consistent with the slow decay in
star formation thought to be acting. It may be that, in some cases,
the transformation to S0 morphology occurs {\it before} the final
cessation of star formation.
MS~0451 passive spirals, in contrast, have an \oii \ equivalent width
consistent with zero, as do their S0 counterparts. Again, this is
consistent with the rapid truncation of star formation in MS~0451
passive spirals, followed by an almost simultaneous transformation
into S0 morphology. The morphological transformation cannot
be significantly delayed after the halt in star formation, because
the S0s' low $D_n(4000)$ and strong H$\delta_A$ strictly limit the
time since last star formation to  $<100$~Myr, according to the model
tracks in Figures~\ref{hd_uvnew} and \ref{dn4000_oii}.

Secondly, the H$\delta_A$ values of passive spirals in each cluster are
similar to those of their counterpart young S0s.
In Cl~0024, passive spirals exhibit weaker H$\delta_A$ than normal star
forming spirals--H$\delta_A=1.8\pm0.2\mbox{\AA}$ in the coadded
spectrum compared to a median H$\delta_A=3.9\pm0.2\mbox{\AA}$ for
star-forming spirals \citep[see also][]{moran06}. The
coadded spectrum of young S0s exhibits a
similarly moderate H$\delta_A=1.2\pm0.2\mbox{\AA}$, 
between the typical values for spirals and for ellipticals. 
In MS~0451, passive spirals (H$\delta_A=3.0\pm0.2\mbox{\AA}$) and young
S0s (H$\delta_A=2.2\pm0.2\mbox{\AA}$) exhibit H$\delta_A$ strengths that 
also mirror each other closely. Yet they exhibit overall higher 
strengths than in Cl~0024, another indication that the halt in star 
formation was quite recent for galaxies in MS~0451.

\begin{figure*}[th]
\centering
\includegraphics[width=\columnwidth]{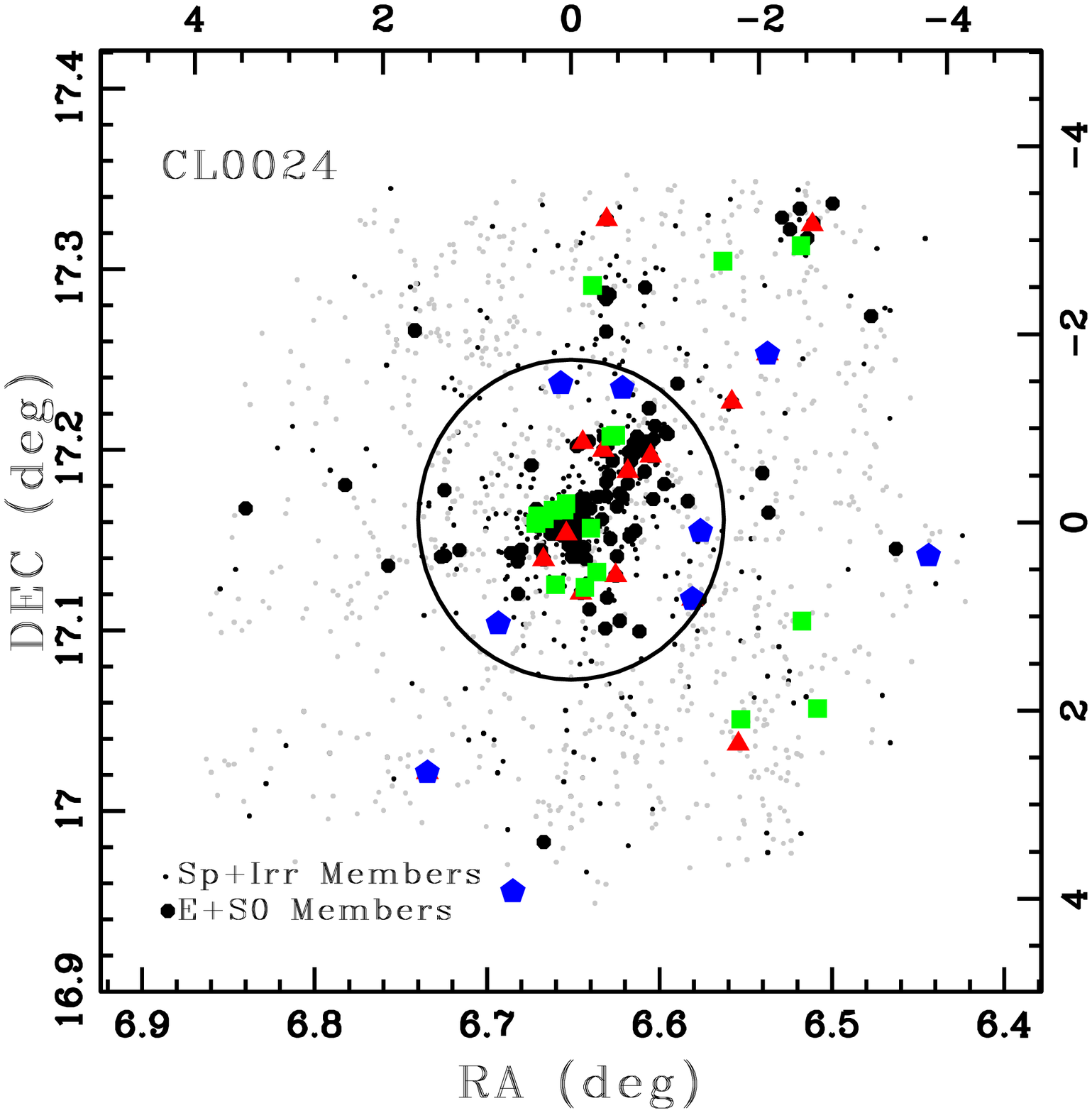}
\includegraphics[width=\columnwidth]{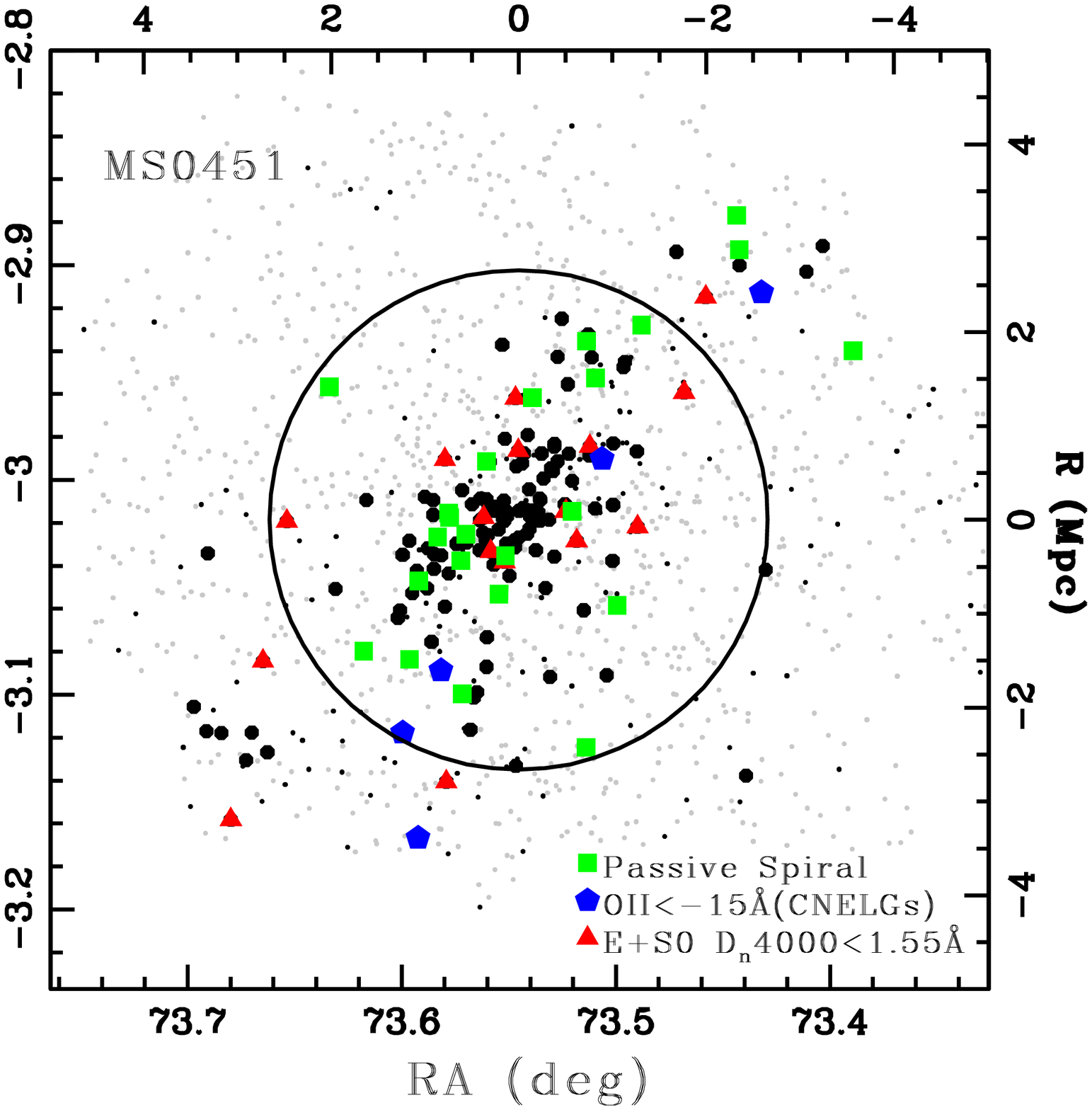}
\caption{\label{spatial_dist}  Distribution of galaxies in the field
  of Cl~0024 (Left), and MS~0451 (Right).
  Galaxies showing spectroscopic signs of recent
  evolution are marked. Spectroscopically confirmed cluster members
  are marked with large (E+S0s) and small (Spirals) black dots. Objects
  with spectroscopically determined redshifts outside the cluster
  are marked as grey dots. Blue pentagons are compact narrow emission
  line galaxies (\oii$<-15\mbox{\AA}$, and to $M_V<-18.0$). 
  Solid green squares are passive spirals.
  Filled red triangles mark early types with $D_n(4000)<1.55$.
 Top and right axes show
  projected radius from the cluster center, in Mpc. Large circle
 indicated the virial radius in each cluster. Passive spirals and
 young S0s are found even in the outskirts of both clusters, perhaps
 residing in groups.}
\end{figure*}

The close similarities between passive spirals and young S0s {\it within}
each cluster, coupled with the clear differences in populations {\it
  between} the two clusters strongly argues in favor of a model where
passive spirals and young S0s represent an evolutionary sequence. 
At the same time, we have not yet fully explored possible explanations
for why this evolutionary sequence behaves differently in the two
clusters. In the next section, we begin consideration of this issue by
examining the local environments in which passive spirals and young S0s
are found.

\section{THE ENVIRONMENTS OF PASSIVE SPIRALS AND YOUNG S0S}

So far, we have treated the populations of passive spirals and young
S0s largely as a uniform population within each cluster. The
reality, however, is more complex. While the passive spirals in 
MS~0451 were largely found to have
different star formation histories than Cl~0024 passive spirals, the
segregation is imprecise: MS~0451 contains a proportion of 
UV-detected spirals ($\sim1/3$) 
consistent with a slow decline in star formation,
while Cl~0024 conversely contains a similar fraction 
of passive spirals whose star
formation may have been truncated more rapidly. To help disentangle
this puzzle, we focus now more closely upon the local environments in
which the passive spirals and their remnant S0s are found.

We start by examining the spatial distribution of transition galaxies
across both clusters. 
In Figure~\ref{spatial_dist}, we plot the positions of all passive
spirals (green squares) and young S0s (red triangles). We also plot
the populations of compact narrow emission line galaxies (CNELGs)
found in each cluster (blue pentagons), which will be discussed
further in \S7.

The first striking feature to note in Figure~\ref{spatial_dist} is
that passive spirals and young S0s occur both in the cluster core and
at higher radius, even beyond $R_V$, with no obvious segregation between
the passive spirals and S0s.  Secondly, in MS~0451 both passive spirals and 
young S0s appear to be more spread out across the cluster than 
in Cl~0024, where more than half are located within 1~Mpc of the
cluster center. 
Both features reveal important clues to the physical processes at work.

Careful inspection of those passive spirals found beyond the cluster
cores gives the impression that many have other cluster members
nearby. 
Could passive spirals be preferentially found in infalling groups?
Some evidence for passive spirals in groups at these redshifts has
already be noted \citep{jeltema07}.
To evaluate quantitatively this possibility, we calculate the
projected local densities of these passive spirals, using the 10
nearest neighbors method, following Paper I and \citet{d97}. Considering only
galaxies outside of the cluster cores ($R>1$~Mpc in Cl~0024 and
$R>1.5$~Mpc in MS~0451, $\sim50\%$ of $R_V$), 
we find that passive spirals are found at a
median local density of $\Sigma=36\pm5$~Mpc$^{-2}$ in MS~0451 and
$\Sigma=51\pm15$~Mpc$^{-2}$ in Cl~0024, both somewhat higher than
the density of a typical infalling spiral of
$\Sigma=28\pm3$~Mpc$^{-2}$ across both clusters.

Passive spirals, therefore, seem to be forming both within infalling groups
and closer to the cluster core--two distinct regions where the
dominant physical forces acting on galaxies are likely to be quite
different. Remarkably, we find that four out of the five UV-detected
passive spirals in MS~0451 reside in groups outside the cluster core.
Recalling from Figure~\ref{hd_uvnew} that these UV-detected passive
spirals closely follow a `starvation-like' gradual
cessation of star formation, we conclude that {\it nearly all} 
passive spirals 
within infalling groups are experiencing a cessation of star formation
spread over a $\sim1$~Gyr timescale.

In contrast, virtually all MS~0451 passive spirals within 1.5~Mpc of the
cluster center have undergone a rapid truncation in star
formation. In Cl~0024 it appears that rapid truncation of star
formation can also occur, but only extremely close to the cluster core: $4/6$ 
passive spirals without UV detection are located within 300~kpc of the
Cl~0024 center.

While passive spirals are abundant in the region around the
MS~0451 center, there is a hole of $\sim600$~kpc radius, within which we
observe almost no spirals of any type. 
Unlike in Cl~0024, where
spirals are observed even within $\sim100$~kpc of the central galaxies, no
spirals survive to reach the center of MS~0451. 
Even accounting for
the larger virial radius of MS~0451, this spiral-free zone 
is significantly larger than that seen in Cl~0024: within $0.2 R_V$
($\sim530$ kpc in MS~0451 and $\sim340$ kpc in Cl~0024),
only $4\%\pm3\%$ of MS~0451 members in our sample are spirals 
($<2\%$ passive), compared to $42\%\pm7\%$ ($9\%\pm3\%$ passive) 
in Cl~0024. 
As the time required for a galaxy to travel $\sim600$kpc 
across the core of MS~0451 is only 0.4~Gyrs, it appears that 
spirals in this environment 
must indeed be converted to early morphology quite rapidly.

\begin{figure}
\centering
\includegraphics[width=3.25in]{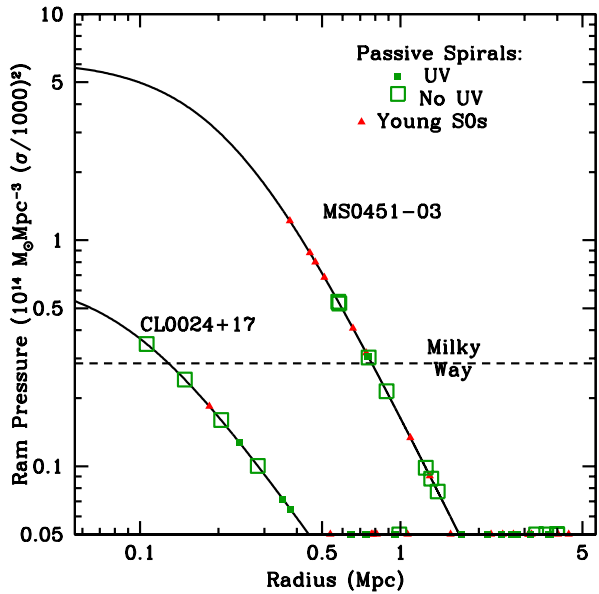}
\caption{\label{rampressure} Strength of ram pressure as a
  function of radius, for both Cl~0024 and MS~0451. Solid lines
  indicate ram pressure $P(R)$ as described in the text. Dotted line
  indicates the pressure required to strip a spiral with
  characteristics of the Milky Way \citep[Paper I][]{gunn72}. Overplotted
  on each track are the radial positions of passive spirals
  (UV-detected and undetected) and young S0s in that cluster. 
  Where the ram pressure
  falls below the lower bound of the plot, we mark the positions of
  galaxies along the lower edge of the plot. Most UV-undetected
  passive spirals are found in regimes of high ram pressure, while no
  passive spirals survive in regimes where ram pressure is more
  than a few times that required to strip a Milky Way-like spiral.}
\end{figure}

Both the central hole in
MS~0451 and the central concentration of UV-undetected passive spirals 
in Cl~0024 can be better understood by considering the local ICM
densities of the transition galaxies in each cluster. 
In Figure~\ref{rampressure}, we make use of
{\it Chandra} X-ray data for each cluster to plot the
expected strength of ram-pressure as a function of radius for Cl~0024
and MS~0451. Each track is generated by calculating the gas density
$\rho(R)$ from the best-fit isothermal
$\beta$-model from \citet{donahue03} and \citet{ota04}, for MS~0451
and Cl~0024, respectively. We then estimate ram pressure $P=\rho v^2$ by
adopting each cluster's line of sight velocity dispersion,
$\sigma$, as the characteristic velocity of a galaxy in that cluster.

Figure~\ref{rampressure} reveals that virtually all UV-undetected
passive spirals are found in regimes where the ram pressure is
significant ($>20\%$ of that required to strip a Milky Way analogue),
while UV-detected passive spirals are largely confined to regions of
lower gas density. While the corresponding radial ranges differ 
between Cl~0024 and
MS~0451, the UV-undetected passive spirals span a nearly-identical
range of ram pressure strengths in each cluster. This argues strongly
in favor of ram pressure stripping as the mechanism responsible for
the UV-undetected passive spirals.  We note that no
spirals survive in environments where the ram pressure is more than a few
times that required to strip the Milky Way, and a natural consequence of
this limit is that no spirals in MS~0451 survive to reach $R<600$~kpc
(though, as expected, several young S0s appear at slightly lower radii).
In \S7.2, we will discuss in more detail the relation between this 
presumed ram pressure stripping and
other mechanisms that may be acting to transform morphology.

\section{PHYSICAL PROCESSES DRIVING THE TRANSFORMATION}

So far, we have shown in \S4 and \S5 that the large population of 
passive spiral galaxies in
both Cl~0024 and MS~0451 are transforming into S0s on a variety of
timescales. Building on this, in \S6, we have successfully 
matched the timescales of transformation with specific 
environments within each cluster.
Consequently, we are now able to construct a road map 
describing the sites
and timescales of transformations across both clusters. 

To summarize, we find that 
some passive spirals are generated
within infalling groups, experiencing a gradual decline in star
formation rate and eventual transformation to S0 morphology.
In the central regions of both clusters, a high proportion of spirals
are seen to be passive. In Cl~0024, star formation within these 
spirals continues to decay at a relaxed pace, with few signs of rapid
interaction with the cluster environment except near the very center
where the effects of ram pressure begin to be important. 
In contrast, passive spirals
all across the core of MS~0451 have had their star formation truncated
rapidly, with a subsequent rapid transformation into S0 morphology.

In this section, we use this road map to consider the constraints 
that we can place on the
physical processes operating in each of these three regimes: groups in
the cluster outskirts, the Cl~0024 core, and the MS~0451 core.

\subsection{The Cluster Outskirts}

In the hierarchical 
assembly of clusters, groups infalling 
from the cluster outskirts can be thought of as the
building blocks. Paper I argued that the existence of the 
morphology--density relation implies that a galaxy's evolutionary
history is tightly linked to its group's history, 
until the group is absorbed into the cluster. 
Furthermore, the observation that mass and light trace each other
tightly on large to small scales (Paper II) suggests that groups of
roughly constant mass to light ratio--rather than individual
galaxies--coalesce to build up the cluster. 

The `preprocessing' of galaxies within these infalling groups is a crucial
feature of evolution within both clusters, as here we are probing galaxies
in a key density range at a key point in time:  at
$z>0.5$, observations indicate that the early type
fraction is growing only in the cores of clusters, 
while below $z\sim0.5$ evolution begins
to accelerate in lower density regions \citep{smith05b, postman05}. As
such, gaining an understanding of the physical processes driving 
galaxy transformation within the infall regions of 
clusters at $z\sim0.5$ can shed light on
the observed evolution in the morphology--density relation.

By virtue of the fact that passive spirals in the cluster outskirts
are preferentially found in groups, we do not expect that the clusters'
ICM plays a direct role in suppressing star formation. Even in MS~0451,
the force of the ICM should not be significant beyond the virial
radius (Figure~\ref{mechanisms}). And while we had previously argued
in favor of starvation by the ICM \citep{moran06}, 
which can be effective at large
cluster radius, there is little reason to expect
it to act on galaxies in groups more effectively than isolated
spirals.

However, we cannot rule out the presence of gaseous interactions {\it
  within} groups. Indeed, X-ray observations of groups in
the local universe reveal that they often harbor an intra-group medium
\citep{mulchaey00}, and simulations by \citet{bekki02} suggest that 
the starvation mechanism can operate similarly within large groups as
it would in the overall cluster \citep[see also][]{hester06,
  fujita04}. 
Intriguingly, X-ray bright groups
seem to contain a higher early-type fraction than 
X-ray undetected groups \citep{zabludoff98}, consistent with the notion that
  the gas plays a role in shaping galaxy morphology. 

The predicted timescale for cessation of star formation under
starvation is consistent with our observed decline over $0.5-2$~Gyr
\citep{fujita04}; in simulations by \citet{okamoto03}, 
a timescale of 1--2~Gyr for the halt in star formation after gas 
removal provided a good fit to observations.
We emphasize, however, that the $\sim1$~Gyr decline in star formation 
represents only one observable phase of the predicted total 
timescale for conversion of
a spiral to an S0 via starvation \citep[$\ge3$~Gyr,][]{bekki02, boselli06b}.
The slow action of starvation would preserve recognizable 
spiral structure for over 1~Gyr after star formation stops, and
we have shown in \citet{moran06} that these lifetimes for each phase
are consistent with the hypothesis that all such passive 
spirals will become S0s by $z=0$.

Yet direct observations of the intra-group gas are challenging 
at these redshifts, especially for small groups \citep[but see][]{jeltema07},
and other mechanisms may also be able to reproduce the observed timescale
for the cessations of star formation. In addition, some models suggest
that modification of star formation rates due to starvation only begin
to be visible many Gyr after the halo is stripped \citep{boselli06b},
and so the effectiveness of starvation in groups would depend in some
sense on the dynamical age of each group, which is difficult to quantify.
Gentle gas stripping or starvation by a diffuse intra-group
medium therefore remains an attractive, but difficult to test, possibility.
 
Alternatively, the increased frequency of galaxy--galaxy interactions
among members of a group, compared to the interaction rate between
isolated galaxies, could be driving the creation of passive spirals in
the cluster outskirts. We note that, at the group scale, the effects
of impulsive galaxy--galaxy interactions are difficult to
separate from galaxy--galaxy tidal interactions that act over a longer 
timescale, and so we consider them together here. 
While galaxy--galaxy harassment at the typical
velocity dispersions of bound groups has not been widely studied,
there are some advantages to this explanation. First, on a larger
cluster scale, simulations indicate that harassment of massive cluster
spirals results in gas being funneled to the central bulge
\citep{moore99}, providing a natural mechanism to accelerate star
formation and deplete a spiral's gas supply; a similar gas funneling
effect due to tidal effects during group preprocessing
has also been proposed \citep[e.g.,][]{mihos04}. 

While simulations would
be necessary to test the hypothesis, one could imagine a 
`gentle harassment' mechanism that induces a transport of gas to the
galaxy center without immediately destroying the spiral arms. 
A slow group-scale harassment 
could possibly reproduce the timescales for quenching of star formation 
indicated by the `starvation' model tracks in Figure~\ref{hd_uvnew}.
We note that the timescale for final conversion to S0 morphology may be
different from the timescale under starvation, and could provide 
an important discriminator between the
two possibilities in the future. 

Depending on the rate of gas
transport, it is possible that a brief increase in star 
formation rate could be detected before the final quenching 
\citep{fujita98, kodama01}, and this could provide another way to 
distinguish between this mechanism and starvation. However, within 
the uncertainties, we cannot confirm any small starbursts among 
galaxies within infalling
groups in our sample. Dividing our total sample of 
star-forming galaxies by radius into three equally-sized bins, we do not
see evidence for an enhancement of either the mean or median \oii \ of a
star forming galaxy in any of the three radial zones. 
Nor do we see any change in the fraction of star-forming galaxies with 
strong ($<-20\mbox{\AA}$) \oii.
However, as the expected lifetime for a burst is short
\citep{fujita04}, we cannot rule out the presence of small bursts that
change the median specific star formation rate by less than 20\%.

A definitive distinction between gas-related processes and galaxy
interactions therefore remains elusive. Importantly, however,
we have already established that galaxy--galaxy harassment
is affecting spirals across both clusters, and at a level that does
not destroy their spiral structure: \citet{moran07a} found that
the high scatter in the cluster Tully-Fisher relation compared to the
field was most readily understood if galaxy harassment is acting to
perturb the kinematics of spirals. \citet{moran07a} 
also observed a strange deficit in the cluster outskirts 
of spirals with high central mass concentrations; one possible explanation 
is that these galaxies have dropped out of our emission line sample as 
they become passive due to the aforementioned funneling of gas to 
their centers.

While hardly a settled issue, we expect that the action of
harassment is more likely to be at the root of the galaxy
evolution in infalling groups, as the mechanism is already known to
be acting in Cl~0024 and MS~0451. Yet some contribution from the
intra-group gas is also possible, and the two effects could act in combination.
Future study of the bulges of spirals and S0s could help distinguish
between the mechanisms by establishing whether significant mass has 
been funneled to centers of galaxies during their evolution, as
predicted in the case of harassment.

\subsection{The Cluster Cores}

We turn now to the cores of both clusters, where spirals in MS~0451
appear to have their star formation quenched at a more rapid pace than
in Cl~0024.
The most obvious explanation for the accelerated truncation 
of star formation within the central region of MS~0451 is, of course,
its much denser ICM. As we saw in Figure~\ref{rampressure}, rapid
truncation of star formation appears to occur in both clusters only in
regimes where the ram pressure is high. Due to the dense ICM in
MS~0451, such an environment is much more widespread in this cluster,
and so, unlike in Cl~0024, most of the passive spirals in its core are
experiencing a rapid truncation of star formation.

Ram pressure stripping, in regimes where it is
effective, can strip an entire spiral disk in less than 100~Myr
\citep{quilis00, fujita99}. Detailed analysis of local analogs of passive 
spirals thought to be undergoing ram-pressure stripping suggest a 
timescale of $\sim300$~Myr for the process to produce a galaxy similar to the
UV-undetected passive spirals we observe \citep{boselli06b,
  cortese07}.  
Both of these are consistent with the maximum timescale of
400~Myr implied by the 600~kpc hole devoid of spirals at the center of
MS~0451. Further, there is some evidence that
post-starburst/post-starforming galaxies in the Coma cluster--similar
to, but fainter than those in MS~0451--are generated via
ram pressure stripping \citep{poggianti04}.

The fact that we observe passive spirals at all implies that
the transformation to S0 is not precisely simultaneous with the halt
in star formation \citep{poggianti99}. However, the `truncated' star formation history 
models presented in Figure~\ref{hd_uvnew} suggest that $D_n(4000)$
strength will increase at a fast rate after truncation, such that
we must be seeing young S0s within 100~Myr of the halt in star formation.
The similar fraction of galaxies observed in both the passive
spiral phase and the young S0 phase support the idea that the
lifetimes for each phase are comparable, at about 100~Myr.

Yet there are several remaining uncertainties in this simple picture.
Ram pressure stripping alone does not alter disk kinematics and spiral
structure on such a short timescale, so there must be an additional
mechanism to speed the conversion. In the cluster cores, harassment
and/or tidal interactions could provide this additional impetus
\citep[e.g.,][]{cortese07, mihos04}. In
Cl~0024, in the absence of rapid truncation by the ICM, a similar
harassment or tidal process must be at work. However, the complex
substructure of Cl~0024, and particularly the effects of the recent
cluster collision, may generate importance differences between the two
clusters. 

In order to better constrain the mechanisms affecting passive spirals in
the cores of Cl~0024 and MS~0451, we are aided by two supplementary 
observations, one of which illuminates the action of harassment, with 
the other providing further constraints on ICM interactions--particularly the
hypothesis that shocks in the ICM could be important.

\subsubsection{Compact Emission-line Ellipticals}
As discussed in \S2, we found in Cl~0024 a population of relatively 
faint elliptical galaxies with strong emission,
concentrated in a narrow range in radius close to the virial
radius.   
We cast doubt on the idea that major mergers were 
the cause of these apparent starbursts,
and we speculated that another rapidly acting physical interaction
must be triggering these bursts of activity.  
Strong shocks in the ICM of Cl~0024 could easily have been
generated \citep{roettiger96} during the cluster-subcluster 
merger \citep{czoske02}, and these shocks could be responsible 
for the triggered starbursts/AGN we observe. 

If this were the case, we hypothesized that we would not
see such bursts in MS~0451, as its smoother large scale structure makes
it less likely that such shocks would be generated. Referring back to
Figure~\ref{spatial_dist}, however, we see that objects of this type
are indeed found in the outskirts of both clusters. While only half as
many are identified in MS~0451 as in Cl~0024, this detection rate is
consistent with the lower spectroscopic completeness of our MS~0451
campaign for galaxies at these low luminosities: in
MS~0451, 10\% of detected cluster members are at $M_V>=19.6$, while in
Cl~0024, the fraction is 20\%.

The properties of these objects--compact size,
\oii$<-15\mbox{\AA}$, and typical luminosities $M_V=-19.2$,
associate them with a class of objects dubbed Compact Narrow Emission Line
Galaxies (CNELGs) \citep{koo95}, or their more luminous cousins the
Luminous Blue Compact Galaxies (LBCGs)
\citep[e.g.,][]{noeske06}. Analysis by \citet{rawat07} of deep ACS
imaging of LBCGs reveals that more than $1/3$ show signs of a recent
merger. Their positions outside of dense regions in both MS~0451 and
Cl~0024 support the notion that these galaxies may simply represent
the remnants of a merger between small, gas-rich galaxies. 

In this interpretation, two important constraints can be put on the
physical processes acting on passive spirals. First, we have
eliminated the possibility that strong shocks in the Cl~0024 ICM have a
significant effect, even on low mass galaxies. Secondly, the minimum
cluster radii at which these CNELGs/LBCGs are found identifies
empirically the point at which merging becomes impossible and
harassment must begin to dominate any galaxy-galaxy interaction.

In Cl~0024, this occurs nearly at the virial radius, while in MS~0451
it occurs at a similar projected radius of 1.5~Mpc. We expect, then,
that harassment within 1.5~Mpc should scale in strength similarly
between the two clusters. If harassment is responsible for the slow
conversion of spirals to S0 in the Cl~0024 core, it should also be
acting on MS~0451 spirals with similar strength.

\subsubsection{Fundamental Plane}

\begin{figure*}[t]
\centering
\includegraphics[width=3.25in]{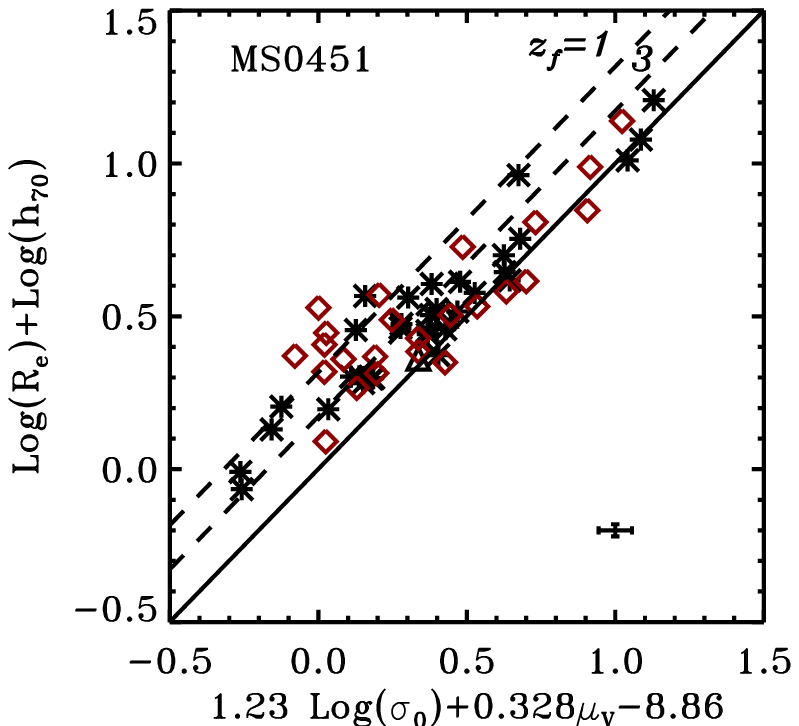}
\includegraphics[width=3.25in]{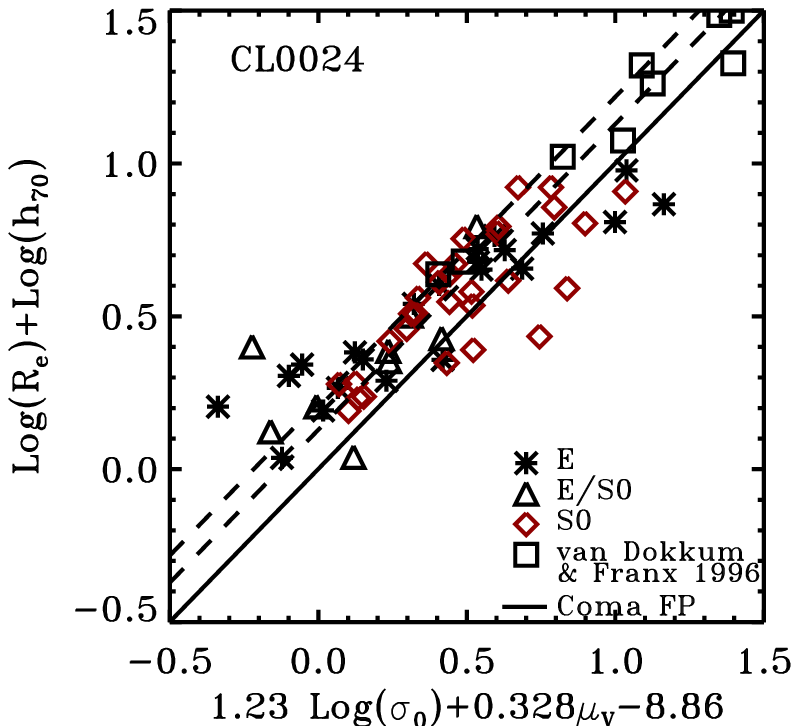}
\caption{\label{fp}The Fundamental Plane of E+S0s in MS~0451, left,
  and Cl~0024, right. Symbols represent different morphologies, as
  indicated in the legend. Solid line is the local FP for the Coma
  cluster, adapted from Lucey et al. (1991). Dashed lines indicate the
  expected positions for early types that formed in a single burst at
  $z=1$ or $z=3$, calculated according to the models of \citet{bc03}. 
  A typical errorbar is shown in the lower right of
  the MS~0451 panel.}
\end{figure*}

An analysis of the Fundamental Plane can provide sensitive constraints
on the star formation and kinematic histories of E+S0s (Paper III).
In Cl~0024, we observed that elliptical and S0 galaxies form a
clear FP (Figure~\ref{fp}), yet they exhibit a high scatter, equivalent
to a spread of $40\%$ in mass to light ratio ($M/L_V$).\footnote{As discussed in Paper III, a galaxy's deviation from the FP can
be thought of as a change in mass to light ratio, via the relation
$\Delta \log (M/L_V)=\Delta \gamma/(2.5\beta)$, where $\beta$ is the
slope of the FP and $\Delta \gamma$ is each galaxy's deviation from
the cluster's overall FP intercept $\gamma$.} 
Upon closer inspection, this high scatter appeared to occur only among
galaxies within 1~Mpc of the cluster core. 
We suggested in Paper III that the enhanced scatter in the
cluster core may somehow be related to the recent cluster-subcluster 
merger in Cl~0024 \citep{czoske02}. By examining the FP and its
residuals for MS~0451 as well, we can test the hypothesis that galaxy
interactions in the core of Cl~0024 are enhanced over that expected for
a relaxed cluster.

In constructing the FP for both clusters, surface brightness 
($\mu_V$) and effective radius ($R_e$)
are determined via the GALFIT software \citep{peng02}, which fits a 2D
model to the {\it HST} imaging. We fit models following the 
deVaucoleur's function to each galaxy (see Paper III), and 
apply a k--correction from observed F814W magnitudes
to rest-frame $V$-band surface brightness, $\mu_V$, as discussed in \S3.

The resulting FP for our sample of 60 E+S0 galaxies in MS~0451 is
shown in Figure~\ref{fp}, alongside the previously published FP for
Cl~0024. For both Cl~0024 and MS~0451, we overplot the
local FP from the Coma cluster \citep{lucey91} as a solid line,
adapted to our chosen cosmology. In addition, we plot 
two parallel dashed lines indicating the galaxies' expected FP intercept
if they had formed in a single burst at $z_f=1$ or $z_f=3$, calculated
using the \citet{bc03} models. It is clear
that the E+S0 populations in both clusters have largely formed their
stars at $z>1$. Nevertheless, the high scatter in the Cl~0024 FP
also seems to be present in MS~0451. Limiting
our sample to those galaxies with $\sigma>100$~km~s$^{-1}$ as in
Paper III, the overall
scatter is $0.18\pm0.02$, expressed in terms of a spread in $Log(M/L_V)$,
compared to $0.16\pm0.02$ for Cl~0024.

A small portion of this large scatter in $M/L_V$ can be attributed to the
higher redshift of MS~0451, where an equivalent spread in galaxy
formation ages translates into a larger FP scatter at $z=0.54$ than at
$z=0.4$. In Figure~\ref{fp}, this effect is most easily seen by
noting the larger separation between the $z_f=1$ and $z_f=3$ model
lines in the MS~0451 panel, compared to the Cl~0024 panel. If MS~0451
early-types were to passively evolve between $z=0.54$ and $z=0.4$, the
single stellar population (SSP) models of \citet{bc03} predict a 
decrease in the FP scatter of only up to $\sim0.04$.

\begin{figure}
\centering
\includegraphics[width=3.25in]{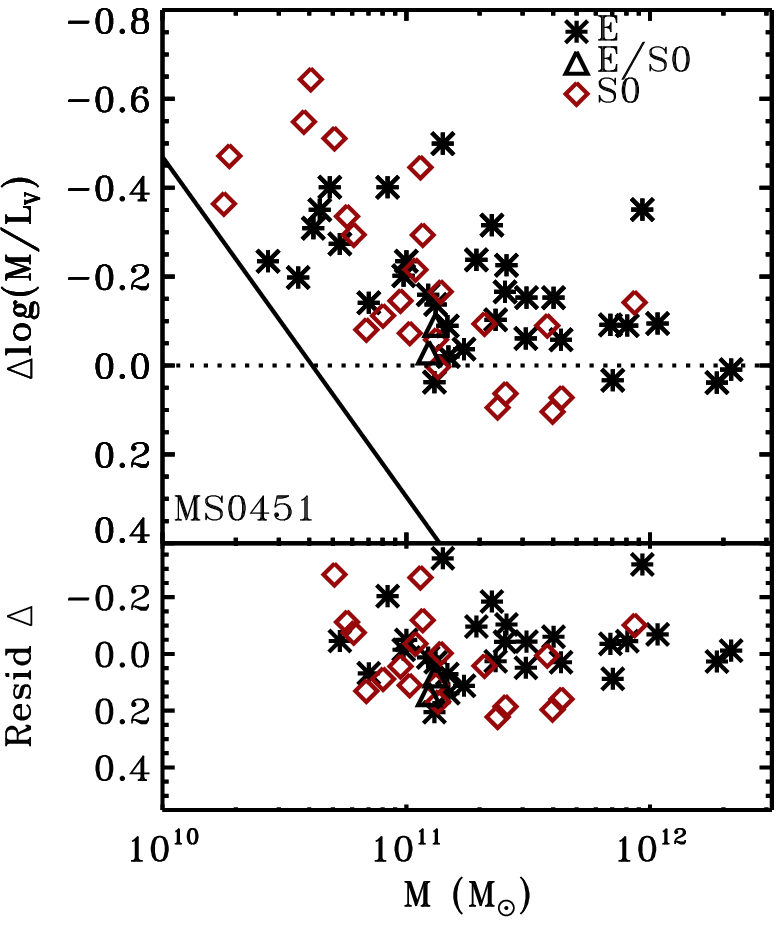}
\includegraphics[width=3.25in]{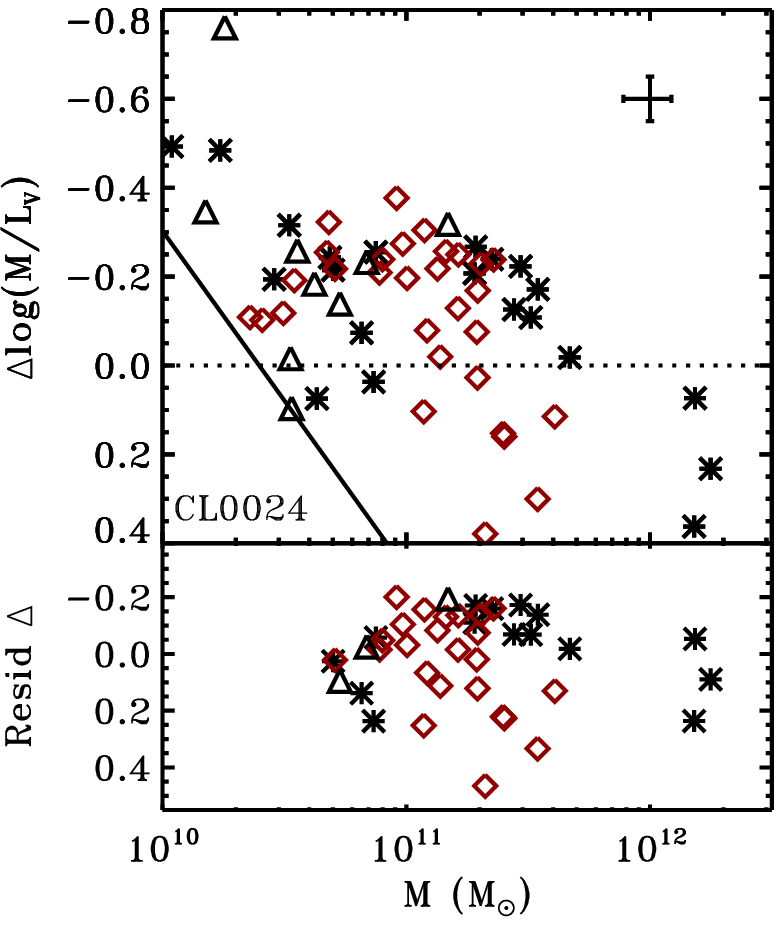}
\caption{\label{deltaml}$\Delta Log(M/L_V)$ versus dynamical mass, M, 
 for MS~0451 (top) and Cl~0024 (bottom). Symbols denote different
 morphologies as indicated on the legend. We are insensitive to
 galaxies in the area to the lower left of
 each solid black line, due to the luminosity limit of our sample. Above,
 $\sim5\times10^{10}$M$_\odot$, we are nearly unbiased. Lower split
 panels show the residuals from fitting and subtracting a straight line
 to galaxies at M$>5\times10^{10}$M$_\odot$. $\Delta
 Log(M/L_V)$ for S0s in Cl~0024 do not form a tight sequence in mass,
 while their counterparts in MS~0451 and the Es in both clusters do
 form a tight sequence. Many Cl~0024 galaxies also occupy unphysical
 locations at $\Delta Log(M/L_V) > 0$.
 }
\end{figure}

This still leaves an unexpectedly high spread in mass to light ratios
for MS~0451 early-types. 
In Figure~\ref{deltaml}, we plot each galaxy's residual from the FP as
a function of its dynamical mass, calculated according to $M=5\sigma^2
R_e/G$ (see Paper III). 
In both clusters, we observe a clear `downsizing' trend that has been
described by several authors \citep[e.g.,][]{tt05b}, 
the tendency for less massive
E+S0s to exhibit lower mass to light ratios, implying younger stellar
populations. 

To judge the effects of `downsizing', we fit a straight line to the
FP residuals as a function of mass, limiting the fit to masses above
$5\times10^{10}$M$_\odot$, where selection effects are minimal. 
In MS~0451, subtracting off this
extra `tilt' to the FP greatly tightens the observed scatter, to
$\Delta Log(M/L_V)=0.12$, a level of scatter consistent with that of
other well-virialized clusters at lower redshift \citep[e.g.,][]{kelson00}. 

Performing the same exercise on Cl~0024, however, results in
no improvement in the FP scatter. Close inspection of the morphologies
of galaxies in Figure~\ref{deltaml} uncovers the reason. While
ellipticals in Cl~0024 display a tight sequence in mass, S0s,
especially at intermediate masses of $\sim10^{11}$M$_\odot$, exhibit a
remarkable scatter in their FP residuals. These residuals do not
appear to correlate with any typical measures of star formation rate
or mass to light ratio, including $D_n(4000)$, H$\delta_A$, or even
broadband optical colors. It seems, then, that the previously
described FP scatter in the core of Cl~0024 can be ascribed almost
entirely to S0s whose properties do not obey any simple trends with
mass or star formation.

Furthermore, a significant number of Cl~0024
galaxies exhibit FP residuals $\Delta Log(M/L_V) > 0$, an unphysical
quantity that either indicates stellar populations older than those of
Coma cluster early types (impossible due to the much younger age of
the universe at $z=0.4$ and the extreme mass to light ratios implied), or else indicates the breakdown of the
fundamental plane for these objects. In contrast, all of the scatter
in MS~0451 early types is in the expected direction, in the sense that
some galaxies have younger stellar populations with $\Delta Log(M/L_V)
< 0$, while the most massive ellipticals and S0s lie neatly on the
local FP.

What could be generating these abnormal galaxies in Cl~0024 but not in MS~0451?
Enhanced levels of galaxy-galaxy interaction due to the complex
substructure in Cl~0024 is the most likely explanation. The
difference between a slow conversion of spirals to S0s (in Cl~0024)
and a rapid truncation of star formation (in MS~0451) could also
affect the positions of S0s on the FP. Future simulations to test this
possibility would be welcome. We note that the actual `young
S0s' are only sparsely represented in Figure~\ref{deltaml}, due to the
smaller sample of galaxies that can be plotted on the FP, and are
not driving the observed difference by themselves.

\bigskip
What, then, are the implications for the physical processes driving
galaxy evolution in the cores of both clusters? In short, we can
conclude that the complex recent assembly history of Cl~0024 
{\it does not} enhance the ability of the ICM to transform galaxies 
via shocks, but
it {\it does} appear to enhance the observed kinematic disturbance of
galaxies within the cluster core. Some mix of galaxy--galaxy
interaction and tidal interaction therefore must dominate in the core
of Cl~0024, driving the steady decline in star formation rate within
its spiral population. Only for galaxies that reach the inner
$\sim300$kpc will ram pressure stripping become important.

In contrast, the quenching of star formation in MS~0451 is
 almost certainly driven by ram pressure from the ICM, probably coupled
 with galaxy and tidal interactions that serve to erase spiral
 structure and complete the morphological conversion. In MS~0451, the
 effects of harassment/tidal interaction must complete the
 morphological conversion more rapidly than in Cl~0024, where the
 stellar populations of passive spirals indicate that they are
 long-lived ($\sim1-2$~Gyr). At the same time, the effects of this
 conversion must somehow decay rapidly enough in MS~0451 S0s
 to be unobserved in the residuals from the FP. Future investigations
 into the effects of harassment on the S0 Fundamental Plane could help
 shed light on this issue.

\section{DISCUSSION AND CONCLUSIONS}

In this paper, we have presented new results from our 
comprehensive comparative survey of two
massive, intermediate redshift galaxy clusters, Cl~$0024+17$ ($z=0.39$) and
MS~$0451-03$ ($z=0.54$). 
We have identified and studied several key classes of transition
objects in the clusters: the passive spirals and the young S0s. 
Through UV imaging and measurements of
spectral line indices, we have concluded that some passive spirals
have experienced a decline in star formation over a $\sim1$~Gyr
timescale, mostly in Cl~0024, while others, mostly in MS~0451, have
experienced a more rapid truncation in star formation.
For the first time, we have been able to conclusively
identify spiral galaxies in the process of transforming into S0
galaxies, by directly linking the passive spirals in each cluster with
their descendant S0s.

Having established that the transformation from spiral to S0 galaxies
is taking place in each cluster, we have leveraged the differences 
between clusters in the timescales and 
spatial location of the conversion process, in order to evaluate the
relative importance of several proposed physical mechanisms that could
be responsible for the transformation.
Combined with other
diagnostics that are sensitive to either ICM-driven galaxy evolution
or galaxy-galaxy interactions--including the residuals from the 
Fundamental Plane and the properties of `signpost' compact emission
line galaxies--we paint a tentative but remarkably self-consistent picture of
galaxy evolution in clusters. 

We find that spiral galaxies within
infalling groups have already begun a slow process of conversion into
S0s primarily via gentle galaxy-galaxy interactions that act to quench
star formation, perhaps aided by interaction with the intra-group
gas. The fates of spirals upon reaching the core of the 
cluster depend heavily on the
cluster ICM, with rapid conversion of all remaining spirals into S0s via
ram-pressure stripping in clusters where the pressure of the ICM is
$\gtrsim20\%$ of that needed to strip a canonical Milky Way-like spiral. In the
presence of a less-dense ICM, the conversion continues at a slower
pace, with galaxy-galaxy interactions continuing to play a role,
perhaps along with `starvation' or gentle stripping by the ICM.

Several authors have raised objections to a scenario where S0s are
created via simple fading of spiral disks. First, S0s are
observed to have higher bulge to disk ratios than their supposed 
spiral progenitors \citep[e.g.,][]{burstein05}, suggesting that
if the spiral to S0 conversion takes place, significant redistribution
of mass or significant new star formation is required
\citep{christlein04, kodama01}. Second, other authors have noted that
the local abundance of S0s is only weakly correlated with environment,
such that processes like ram pressure stripping that act only in
cluster cores cannot be responsible for the entire buildup of
S0s \citep{dressler04, d80}.

We have here demonstrated, however, that simple stripping of gas does {\it not}
build up the entire population of cluster S0s. Rather, a combination
of gas effects and galaxy interactions are responsible, in ratios that
vary widely and depend on both the cluster dynamical state and the
location within the cluster or its outskirts. 
The buildup of bulges through new star formation 
\citep{christlein04} that is expected from harassment-like
encounters could potentially balance out the expected fading of disks
after ram pressure stripping has run its course. It may still be
necessary to invoke significant obscured star formation
to bridge the observed gap between the bulges of spirals and S0s
\citep{kodama01}; the higher detection rate of such systems in Cl~0024
compared to MS~0451 \citep{geach06} may indicate that such obscured 
starbursts are part of the pre-processing that occurs before galaxies encounter
a dense ICM.

Furthermore, the wide
distribution of passive spirals across both clusters, and their
association with groups in the cluster outskirts, indicates that the
resulting S0s will likely be spread across a wide range of
environments. The typical densities of infalling groups at $z\sim0.5$
are likely realized even in isolated groups by $z=0$ \citep{fujita04}, 
and so it seems that, though a variety of mechanisms are responsible, 
passive spirals could be the progenitors of most local S0s.

In \citet{moran06}, a rough assessment of the frequency of passive
spirals and their expected lifetimes in Cl~0024 suggested that passive
spirals could account for the entire buildup of S0s between $z=0.4$
and $z=0$. In MS~0451, we observe a similar number of spirals
currently in the passive phase, but we have shown that the timescale
for conversion is much shorter than in Cl~0024. This
suggests that all existing spirals should be converted to S0 in MS~0451 even
{\it before} $z=0$, and the rate of any future buildup should be
limited by the infall rate of new galaxies. As MS~0451 is a remarkably
well-evolved system at $z=0.54$, comparable in mass to Virgo or
Coma \citep[$6.9\times10^{14} M_\odot$ and $1.1\times10^{15}
M_\odot$, respectively,][]{biviano93,fouque01,geller99}, this is perhaps not surprising. 

It suggests,
however, that there may exist a generic tipping point in the 
evolution of a massive cluster: beyond the threshold in ram pressure
strength that we have identified here (which is reached through
some combination of cluster mass and ICM density), 
spirals are transformed so rapidly upon infall that they 
will be essentially absent. By $z=0$, 
many clusters have likely reached this threshold, as evidenced by
their low spiral fractions \citep{d97, d80}. This sort of dichotomy
between well-evolved and still-assembling clusters could, for example, 
explain why some clusters, like Coma, 
reveal few signs of evolution in their massive galaxies
\citep{poggianti04}, while others,
like Virgo, have a rich population of passive spirals and galaxies
with other signs of recent evolution \citep[e.g.,][]{chung07}.

Regardless, the results presented here indicate that the abundant
cluster spirals found at intermediate redshift, {\it do}, in fact,
transform into the equally abundant S0 population seen today. The
transformation process at $z=0.5$, as we have seen, is beginning to
operate even in groups at the cluster outskirts. As a result, the
conversion of passive spirals to S0s can account for the evolution of the
morphology--density relation at both the cluster and group scale
from intermediate redshift to today.

\acknowledgements

We thank the referee, A. Boselli, for insightful comments, especially
on the various physical processes and their timescales, which were helpful in
revising the discussion of these issues.
SMM would like to thank T. Heckman, G. Kauffmann, S. Yi,
\& members of the {\it GALEX} science
team for valuable discussions. Faint object spectroscopy at Keck Observatory is made possible with
LRIS and DEIMOS thanks to the dedicated efforts of J. Cohen,
P. Amico, S. Faber and G. Wirth. We acknowledge use
of the Gauss-Hermite Pixel Fitting Software developed by R. P.
van der Marel. The analysis pipeline used to reduce the
DEIMOS data was developed at UC Berkeley with support from
NSF grant AST-0071048. TT acknowledges support from the NSF through
CAREER award NSF-0642621, and from the Sloan Foundation through a
Sloan Research Fellowship. RSE acknowledges
financial support from NSF grant AST-0307859 and STScI grants
HST-GO-08559.01-A and HST-GO-09836.01-A. IRS and GPS acknowledge support from
the Royal Society.

\end{document}